\RequirePackage{fix-cm}
\documentclass[twocolumn]{svjour3}          % twocolumn
\smartqed  % flush right qed marks, e.g. at end of proof

\usepackage{amsmath}

\usepackage{graphicx}
\usepackage[T1]{fontenc}
\usepackage[utf8,latin1]{inputenc}
\usepackage{epsfig}
\usepackage{txfonts}
\usepackage{amssymb}
\usepackage{siunitx}
\usepackage{natbib}

\usepackage{url}
\usepackage{color}
\usepackage{sgl-commands}

 \journalname{ISSI Workshop}

\begin{document}
%\title{Chapter 1: A primer on strong gravitational lensing}
\title{Microlensing near macro-caustics}

%\subtitle{}

%\titlerunning{Short form of title}        % if too long for running head

\author{Luke Weisenbach$^{1}$, Timo Anguita$^{2,3}$, Jordi Miralda-Escud\'e$^{4,5,6}\dagger${\thanks{$\dagger$ \email: miralda@icc.ub.edu}}, Masamune Oguri$^{7,8}$, Prasenjit Saha$^{9}$ \& Paul L. Schechter$^{10}$}

%\authorrunning{Short form of author list} % if too long for running head

\institute{$^{1}$ Institute of Cosmology and Gravitation, University of Portsmouth, United Kingdom \\
$^{2}$ Instituto de Astrofisica, Departamento de Ciencias Fisicas, Universidad Andres Bello, Chile\\
$^{3}$ Millennium Institute of Astrophysics, Chile\\
$^{4}$ Institute of Cosmos Sciences, University of Barcelona, Barcelona 08028, Spain \\
$^{5}$ Instituci\'o Catalana de Recerca i Estudis Avan\c cats, Barcelona 08010, Spain \\
$^{6}$ Institut d'Estudis Espacials de Catalunya, Castelldefels 08860, Spain \\
$^{7}$ Center for Frontier Science, Chiba University, 1-33 Yayoi-cho, Inage-ku, Chiba 263-8522, Japan \\
$^{8}$ Department of Physics, Graduate School of Science, Chiba University, 1-33 Yayoi-Cho, Inage-Ku, Chiba 263-8522, Japan\\
$^{9}$ Physik-Institut, University of Zurich, Winterthurerstr.~190, 8057 Zurich, Switzerland\\
$^{10}$ MIT Kavli Institute 37-635, Cambridge MA 02139, USA \\
}

\date{}
% The correct dates will be entered by the editor

\maketitle

\begin{abstract}
Microlensing near macro-caustics is a complex phenomenon in which swarms of micro-images produced by micro-caustics form on both sides of a macro-critical curve. Recent discoveries of highly magnified images of individual stars in massive galaxy cluster lenses, predicted to be formed by these micro-image swarms, have stimulated studies on this topic. In this Chapter, we explore microlensing near
macro-caustics using both simulations and analytic calculations. We show that the mean total magnification of the micro-image swarms follows that of an extended source in the absence of microlensing. Micro-caustics join into a connected network in a region around the macro-critical line of a width proportional to the surface density of microlenses; within this region, the increase of the mean magnification toward the macro-caustic is
driven by the increase of the number of micro-images rather than individual magnifications of micro-images. 
%The microlensing property is sensitive to the mass fraction of microlenses (stars) such that 
The maximum achievable magnification in micro-caustic crossings decreases with the mass fraction in microlenses. We conclude with a review of applications of this microlensing phenomenon, including limits to the fraction of dark matter in compact objects, and searches of Population III stars and dark matter subhalos. We argue that the discovered highly magnified stars at cosmological distances already imply that less than $\sim$ 10\% of the dark matter may be in the form of compact objects with mass above $\sim 10^{-6}\, M_{\odot}$.
% add further keywords with "\and" 
\end{abstract}

\newcommand{\vbeta}{\mbox{\boldmath$\beta$}}
\newcommand{\vtheta}{\mbox{\boldmath$\theta$}}
\newcommand{\valpha}{\mbox{\boldmath$\alpha$}}
\newcommand{\vzero}{\mbox{\boldmath$0$}}
\newcommand{\vepsi}{\mbox{\boldmath$\epsilon$}}
\newcommand{\vecx}{\mbox{\boldmath$x$}}
\newcommand{\half}{\mbox{$\frac12$}}
\newcommand{\onesixth}{\mbox{$\frac16$}}
\newcommand*\diff{\mathop{}\!\mathrm{d}}

\section{Introduction}

%{\em For an earlier version see end of source file.}

\citet{1986ApJ...301..503P} showed how a screen of point mass micro-lenses (stars) of sufficiently high surface density produces an extended swarm of micro-images, the sum of whose fluxes is, on average, 
%roughly 
equal to the flux expected for a point source lensed by a smooth mass distribution with the same surface density \citep[see also][]{1986ApJ...306....2K,2003ApJ...583..575G,2011MNRAS.411.1671S}.

A caustic is a closed one dimensional locus in the source plane of a lens system that maps onto a one dimensional locus in the image plane -- a critical curve -- along which images of point sources undergo infinite magnification in the geometric optics approximation. A physical source has a finite angular size that implies a maximum magnification, which is reached when the source straddles a caustic and gives rise to two merging mirror images straddling the corresponding critical curve. 
%The portion of the source that lies inside the caustic is doubly imaged, while there is no image of the portion outside. 
When a source crosses a caustic, a pair of mirror images either annihilates at or emerges from the critical curve \citep[see e.g.,][]{1992grle.book.....S}.

These two distinct phenomena combine when one has micro-lensing near a caustic, with the emergence or annihilation of a pair of elongated swarms of micro-images at a critical curve.  The combination of these two phenomena was considered by \citet{1990PhDT.......180W} as a possible explanation for BL Lac systems.  The discovery of a bright  transient, thought to be a highly magnified star at redshift 1.5, near the critical curve of the lensing cluster MACS J1149.5+2223 \citep{2018NatAs...2..334K} led to renewed interest, triggering theoretical studies of the ``trains'' of micro-images, or elongated swarms, formed on the two sides of a macro-critical curve \citep{2017ApJ...850...49V,2018ApJ...857...25D,2018PhRvD..97b3518O,2021arXiv210412009D} as well as observational searches of more such events \citep{2018NatAs...2..324R,2019ApJ...881....8C,2019ApJ...880...58K,2020MNRAS.495.3192D}.
  
In what follows, after reviewing the macro-lens model near macro-caustics (Section~\ref{sec:macro}), we briefly review the behavior of extended sources in the vicinity of a caustic (Section~\ref{sec:extend}). This is then combined with the behavior of micro-image swarms in regions of high optical depth discussed in the chapter by \cite{2024SSRv..220...14V} to show how these two phenomena modify the predictions for each phenomenon considered separately and how these can be used to interpret transients near macro-caustics (Section~\ref{sec:microlens}). Finally, we review recent observations of micro-caustic crossing near macro-caustics and discuss some applications of these microlensing phenomena, including the possibility of searching for more exotic substructure (Section~\ref{sec:application}).
  
\section{The macro model}\label{sec:macro}
  
%Since we are interested in 
To study microlensing near macro-caustics, we
%need to 
define a macro mass model (i.e., the lens model without microlenses) that contains a caustic and a critical curve,
%. For our purpose, it is natural to adopt a 
and set the coordinate system origin of the image and source planes at a point on the critical curve and the caustic,
%pass through the origins of 
respectively. The Fermat time-delay surface that satisfies this condition is written as
\begin{equation}
    \tau = \frac{1}{2}(\bang - \tang)^2 - \psi(\tang)~,
\end{equation}
where $\bang$ and $\tang$ are two-dimensional vectors of the angular positions in the source plane and the image plane, respectively, with coordinates $(\beta_1,\beta_2)$ and $(\theta_1,\theta_2)$. The Taylor-expanded deflection potential, up to third order at the origin, is
\begin{equation}
\begin{split}
    \psi(\tang) = & \frac{1}{2}(\psi_{11}\theta_1^2 + 2\psi_{12}\theta_1\theta_2 + \psi_{22}\theta_2^2) \\
    +& \frac{1}{6}(\psi_{111}\theta_1^3 + 3\psi_{112}\theta_1^2 \theta_2 + 3\psi_{122}\theta_1 \theta_2^2 + \psi_{222}\theta_2^3)~.
\end{split}
\label{eq:psi_expand}
\end{equation}
Note that the first order terms of the expansion disappear because of the requirement that the origin of the image plane maps to the origin of the source plane. Denoting convergence at the origin as $\kappa_0$, i.e.,
\begin{equation}
    \psi_{11}+\psi_{22}=2\kappa_0~,
\end{equation}
and using the condition that the critical curve and the caustic pass through the origin, the second derivatives of the deflection potential can be described as 
\begin{equation}
\begin{aligned}
  & \psi_{11} = \kappa_0 + (1-\kappa_0)\cos\omega~,  \\
  & \psi_{22} = \kappa_0 - (1-\kappa_0)\cos\omega~, \\
  & \psi_{12}  = -(1-\kappa_0) \sin\omega~, 
\end{aligned}
\label{eq:psi_second}
\end{equation}
where $\omega$ is an arbitrary constant parameter.

Eqs.~\eqref{eq:psi_expand} and \eqref{eq:psi_second} form a general expression of the lens model near the macro-caustic. We have the additional freedom to rotate the coordinate system to simplify it further. For instance, choosing the coordinate system with $\omega=0$ realizes the locally orthogonal coordinate system around a fold caustic, which is our main interest here. We could also choose to instead rotate the coordinate system to eliminate one of the third derivatives of the deflection potential, although such a transformation makes discussion regarding the direction and extent of the image trains less intuitive. 

It is also helpful to consider more restrictive cases for the microlensing analysis. One possible choice is to make convergence constant across the field, for which we can easily attain microlensing simulations with a constant stellar mass fraction to the total matter density. Since convergence $\kappa$ is computed from Eq.~\eqref{eq:psi_expand} as
\begin{equation}
\kappa=\kappa_0+\frac{1}{2}(\psi_{111}+\psi_{122})\theta_1+\frac{1}{2}(\psi_{112}+\psi_{222})\theta_2~,
\end{equation}
we need $\psi_{111}=-\psi_{122}$ and $\psi_{112}=-\psi_{222}$ to realize constant convergence. Setting $\omega=0$ and denoting $\psi_{111}=-1/d_\parallel$ and $\psi_{112}=-1/d_\perp$, the lens equation becomes
\begin{equation}
\begin{aligned}
\beta_1 = &\frac{\theta_1^2}{2d_\parallel} + \frac{\theta_1 \theta_2}{d_\perp} - \frac{\theta_2^2}{2d_\parallel}~,\\
\beta_2 = &2(1-\kappa_0 )\theta_2+ \frac{\theta_1^2}{2d_\perp} - \frac{\theta_1 \theta_2}{d_\parallel} -\frac{\theta_2^2}{2d_\perp}~,
\end{aligned}
\label{eq:lenseq_1}
\end{equation} 
which, apart from notation, is the same as the macro lens model adopted in e.g., \citet{2021arXiv210412009D}.

The parameters $d_\parallel$ and $d_\perp$ have dimensions of the angular variable $\theta$, and are on the order of the Einstein radius of the macro lens model, which is $\mathcal{O}(1'')$ for galaxy-scale strong lensing and  $\mathcal{O}(10'')$ for cluster-scale strong lensing. In this paper we will use a dimensionless variable defined as the angular coordinates divided by the Einstein radius of each microlens, $\theta_\star$, which is $\theta_\star=\mathcal{O}(10^{-6}{}'')$ for typical stars acting as microlenses at cosmological distances. Thus $d_\parallel$ and $d_\perp$ are on the order of $\sim 10^6-10^7$ in this dimensionless variable.

Alternatively, we consider a restricted case where the critical curve and a train of micro-images (which follows the direction of the principal axis that has zero eigenvalue at the critical curve) are perpendicular to each other. This is realized by setting $\psi_{112}=\psi_{122}=\psi_{222}=0$ along with $\omega=0$. Denoting $\psi_{111}=-1/E$, the lens equation becomes
\begin{equation}
\begin{aligned}
\beta_1 = &\frac{\theta_1^2}{2E}~,\\
\beta_2 = &2(1-\kappa_0 )\theta_2~.
\end{aligned}
\label{eq:lenseq_2}
\end{equation} The variable $E$ again has dimensions of the angle $\theta$, and is on the order of $\sim 10^6-10^7$ in the unit defined by $\theta_\star$.

While this macro lens model is a restricted case, it makes the analysis much easier and hence is instructive; we will use it in Section~\ref{sec:extend}. From our experience in modeling such systems, introducing other degrees of freedom related to further derivatives of the potential introduces modest quantitative changes but not major qualitative differences. In this model, convergence changes near the critical line as
\begin{equation}
\kappa=\kappa_0-\frac{\theta_1}{2E}~.
\end{equation}

\begin{figure}
\centering
\includegraphics[width=.9\linewidth]{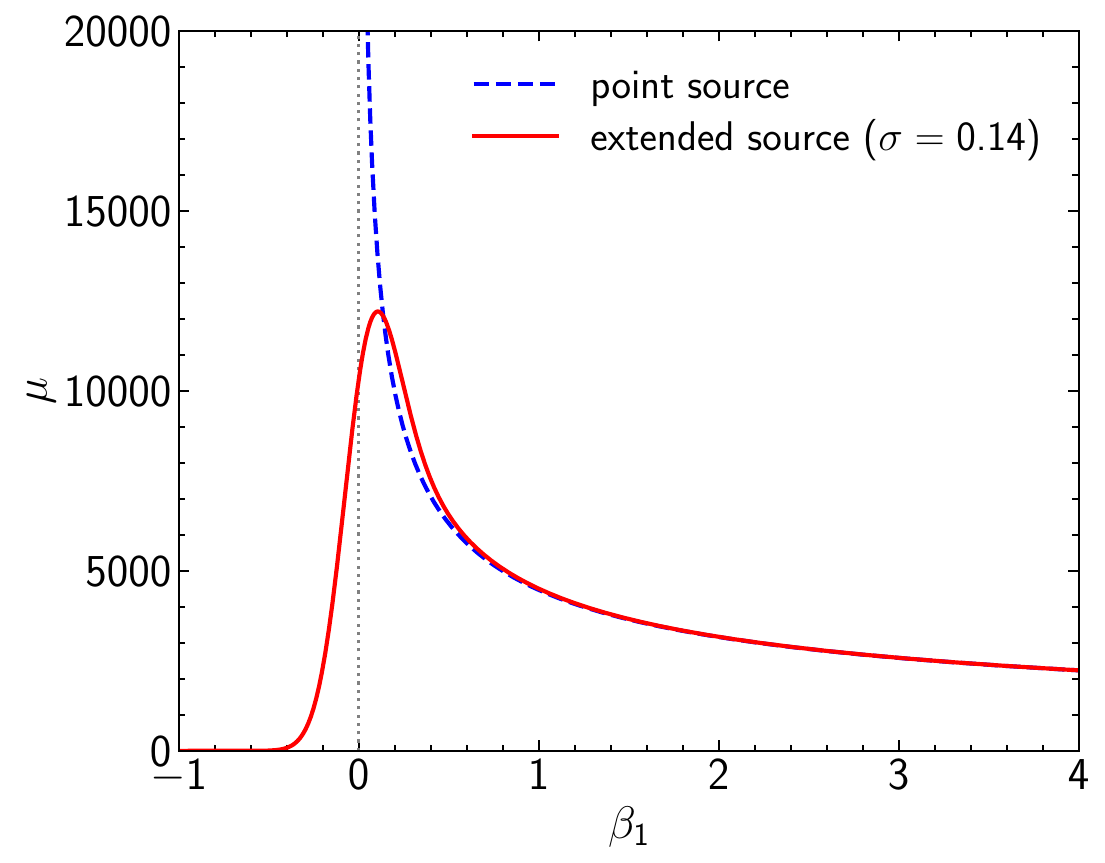}
\caption{Magnification factors as a function of the source position $\beta_1$, both for the point source ({\it dashed}) and the extended source ({\it solid}), in absence of microlensing. We adopt the macro model of Eq.~\eqref{eq:lenseq_2} with $\kappa_0=0.5$ and $E=10^7$. The extended source follows the Gaussian surface brightness distribution with $\sigma=0.14$. }
\label{fig:mag_smooth}
\end{figure}

\section{Extended sources in the vicinity of a caustic}\label{sec:extend}

In this Section, we review the behavior of magnification of extended sources in the vicinity of a macro 
%model 
caustic in the absence of microlensing.
%For this purpose 
We use the simple macro model described by Eq.~\eqref{eq:lenseq_2}. %and start with discussions on magnifications of a point source, ignoring the source size. For 
At an image plane position $\tang=(\theta_1$, $\theta_2$), magnification factors along the $\theta_1$ and $\theta_2$ directions, which we denote $\mu_1$ and $\mu_2$ respectively (equal to the inverse eigenvalues of the magnification matrix), are easily computed as
\begin{equation}
\begin{aligned}
\mu_1 = &\frac{E}{\theta_1}~,\\
\mu_2 = &\frac{1}{2(1-\kappa_0)}~.
\end{aligned}
\end{equation} 
The magnification of a point image at that point is $\mu=\mu_1\mu_2$. A point source at $\bang=(\beta_1$, $\beta_2$) has two images on both sides of the critical curve at $\theta_1=\pm\sqrt{2E\beta_1}$ if and only if $\beta_1>0$. Thus the magnification of these two images of a point source at $\bang$ is
\begin{equation}
  \mu(\bang) = \frac{1}{(1-\kappa_0)}\sqrt{\frac{E}{2\beta_1}}~,
  \label{eq:mu_beta_point}
\end{equation} 
for $\beta_1>0$.
%and $\mu(\bang)=0$ for $\beta_1<0$.

For an extended source with a Gaussian
%. Assuming that the 
surface brightness profile with standard deviation $\sigma$, the magnification factor of the source at $\bang$ is given by
\begin{equation}
  \mu(\bang) = \frac{\sqrt{E}}{2\sqrt{\pi}\sigma(1-\kappa_0)}\int_0^\infty \frac{\diff\beta_1'}{\sqrt{\beta_1'}}\exp\left[-\frac{(\beta_1'-\beta_1)^2}{2\sigma^2}\right]~.
  \label{eq:mu_beta_extend}
\end{equation} 

As an example, Figure~\ref{fig:mag_smooth} shows magnification factors as a function of the source position $\beta_1$, both for the point source and the extended source, for the case $E=10^7$, $\kappa_0=0.5$ and $\sigma=0.14$. The magnification of the point source diverges as the source approaches the caustic at $\beta_1=0$, while the magnification of the extended source saturates at a maximum value that can be estimated as the magnification of the point source (Eq.~\ref{eq:mu_beta_point}) at $\beta_1=\sigma$ \citep[e.g.,][]{1991ApJ...379...94M}, i.e.,
\begin{equation}
\mu_{\rm max}\approx \frac{1}{(1-\kappa_0)}\sqrt{\frac{E}{2\sigma}}~.
\end{equation} 
For the example in Figure~\ref{fig:mag_smooth}, this estimate yields $\mu_{\rm max}\approx 12000$, close to the maximum magnification directly derived from Eq.~\eqref{eq:mu_beta_extend},  $\mu_{\rm max}\approx 12200$. 

%\section{The Corrugated Band of Micro-Critical Lines: Number of Microimages and their characteristic Magnification}
\section{Microlensing near macro-caustics}\label{sec:microlens}

We now describe the impact of microlensing on the magnification of sources.

\begin{figure}
\centering
\includegraphics[width=.9\linewidth]{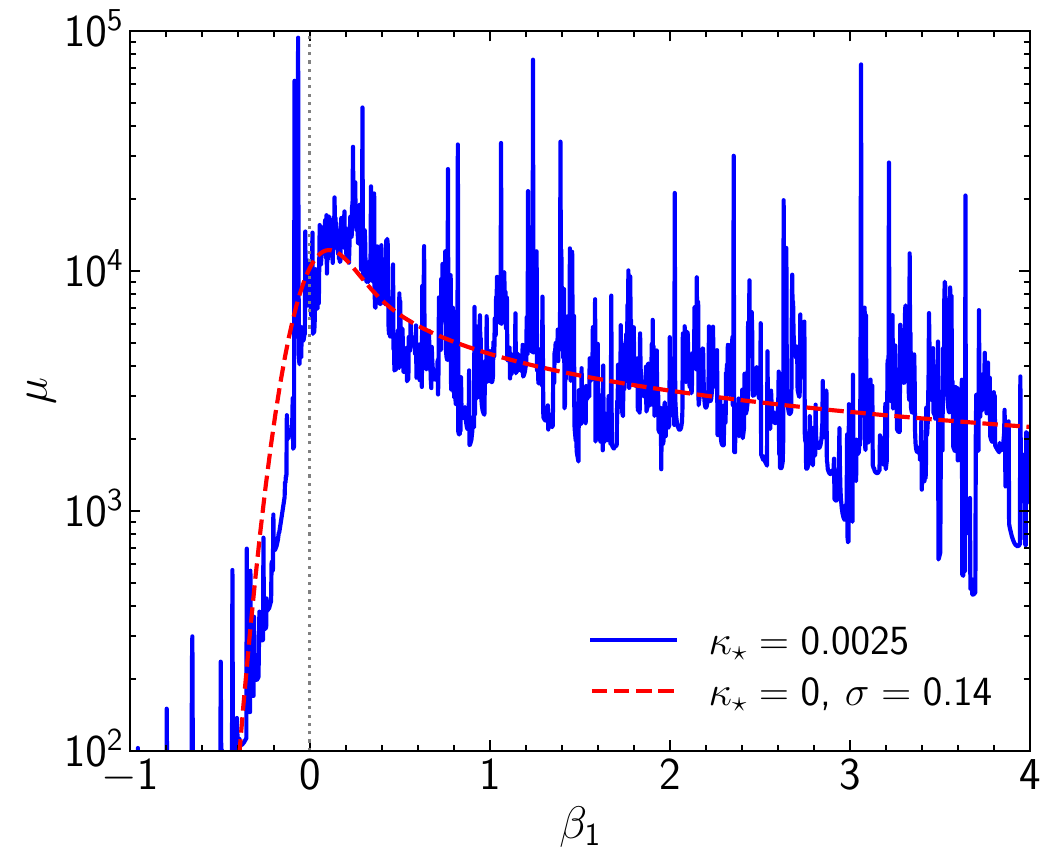}
\caption{Magnification of a point source as a function of source position $\beta_1$ near the macro-caustic in the presence of microlenses ({\it blue solid line}), using the macro model of Eq.~\eqref{eq:lenseq_2} with $\kappa_0=0.5$ and $E=10^7$, and convergence of point masses of $\kappa_\star=0.0025$. For comparison, magnification of an extended source with a Gaussian surface brightness distribution with $\sigma=\sigma_{\valpha}=0.14$ (see also Eq.~\eqref{eq:sigma_alpha}) in the absence of microlenses is plotted by a dashed line.}
\label{fig:train_mag_all}
\end{figure}

\begin{figure}
\centering
\includegraphics[width=.9\linewidth]{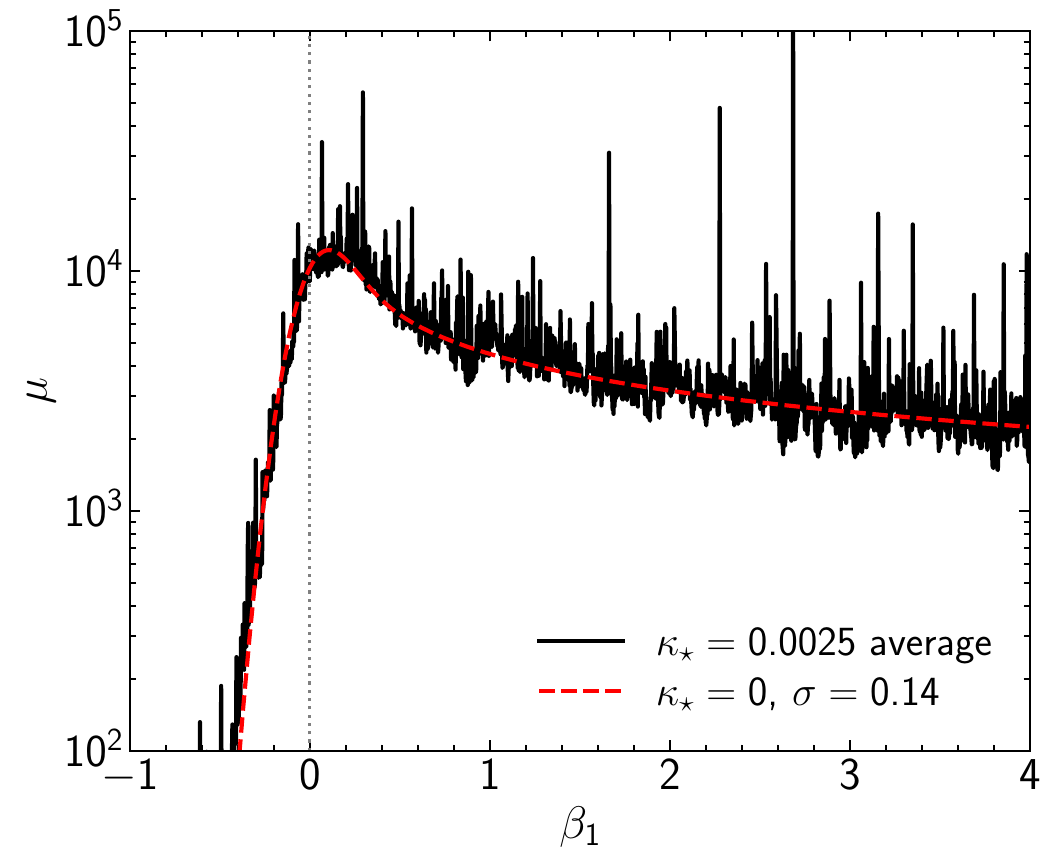}
\caption{Similar to Figure~\ref{fig:train_mag_all}, but averaging over 10 realizations of point mass positions.}
\label{fig:train_mag_all_ave}
\end{figure}

\subsection{Average magnification profile interpreted as an effectively extended source}\label{sec:effective_extend}

As shown in \cite{2003A&A...404...83N} and \cite{2017ApJ...850...49V}, and discussed in the chapter by \cite{2024SSRv..220...14V} for quasars, the total magnification of a source in the presence of microlensing, averaged over all random point mass realizations of the microlenses, is the magnification under the macro-lens model of the source profile convolved with the probability distribution function (PDF) of the microlens deflection angle: 
\begin{equation}
    \langle\mu(\vbeta)\rangle = \int I_{\star}(\vbeta') \mu_{\rm B}(\vbeta + \vbeta')\diff^2 \vbeta' ~,
\end{equation}
where $\vbeta$ is the position of the source, $\vbeta'$ is a dummy variable of integration over the source plane,
%$I_{\valpha}(\vbeta')$
$I_\star(\vbeta')$
is the source profile $I(\vbeta')$ convolved with the PDF of the microlens deflection angle $p(\valpha_\star)$, and $\mu_{\rm B}(\vbeta)$ denotes the magnification factor of the background macro-model (i.e., the mass model after the point mass lenses are smoothed) for a source at $\vbeta$. The mean effect of microlenses is encapsulated in the convolved source profile $I_\star(\vbeta')$.
%As discussed already, 
The PDF of the deflection angle has a bivariate Gaussian core with a width of
\begin{equation}
  \sigma_{\star} = \theta_\star\kappa_\star^{1/2}\big[\ln(2e^{1-\gamma_E}N_\star^{1/2})\big]^{1/2} ~,
    \label{eq:sigma_alpha}
\end{equation}
where $\kappa_\star$ is the convergence of point mass lenses (stars), $\theta_\star$ is the Einstein radius of each point mass, $\gamma_E\approx 0.577$ is the Euler-Mascheroni constant, and $N_\star$ is the number of point masses.

Although the deflection angle distribution $p(\valpha_\star)$ is not precisely a Gaussian, it can be approximated this way whenever the deflection is contributed by many microlenses, according to the central limit theorem (this fails, of course, for the tails of this distribution caused by large deflections near a single microlens).
The behavior of the mean magnification of point sources near a macro-caustic should then follow that of an extended Gaussian source in the absence of microlensing as studied in Section~\ref{sec:extend}. To explicitly check this point, we make numerical simulations of the magnification of a point source in the simple macro model of Eq.~\eqref{eq:lenseq_2}, with $\kappa_0=0.5$ and $E=10^7\theta_\star$, adding randomly distributed microlenses of a single mass corresponding to an Einstein radius $\theta_\star$, within a circle of radius $\theta_{\rm max}=15000\, \theta_\star$. A surface density contributing a convergence $\kappa_\star=0.0025$ 
in this macro model is replaced with microlenses. This stellar mass fraction is similar to the expected value for the first discovery of a micro-caustic crossing event due to stars making up the intracluster light (see Section~\ref{sec:obs}). With the choice of these parameters, from Eq.~\eqref{eq:sigma_alpha} we obtain the effective width of the Gaussian as $\sigma_{\star}\approx 0.14\, \theta_\star$. We solve the lens equation
%in this set-up 
using an adaptive mesh method implemented in the {\tt glafic} software \citep{2010PASJ...62.1017O} together with a hierarchical tree algorithm to speed up calculations of deflection angles from an ensemble of point mass lenses \citep[see e.g.,][]{1999JCoAM.109..353W}. 

Figure~\ref{fig:train_mag_all} shows the result of moving the source along the $\beta_1$ axis, fixing $\beta_2=0$ and computing the total magnification as a function of the source position, or lightcurve. While many sharp peaks corresponding to micro-caustic crossings are present, the mean magnification closely follows the prediction of the total magnification for an extended source without microlenses as studied in Section~\ref{sec:extend}, shown by a dashed line in Figure~\ref{fig:train_mag_all}. Note that the size of the extended source is not fitted to the lightcurve, but determined by Eq.~\eqref{eq:sigma_alpha}. This confirms the average effect of microlensing in macro-caustic crossings can indeed be interpreted as an effectively extended source, as discussed in the literature \citep[e.g.,][]{2021arXiv210412009D}.

The lightcurve in Figure~\ref{fig:train_mag_all} exhibits large fluctuations originating from a random realization of positions of point masses. A better comparison with the model of an effectively extended source is obtained by averaging the calculation of the lightcurve over independent realizations of the point mass positions. The mean of 10 realizations is shown in Figure~\ref{fig:train_mag_all_ave}, indicating a converging match of the lightcurve of an effectively extended source with the numerical result.
%, which confirms the validity of such interpretation.

\begin{figure*}
\centering
\includegraphics[width=.99\linewidth]{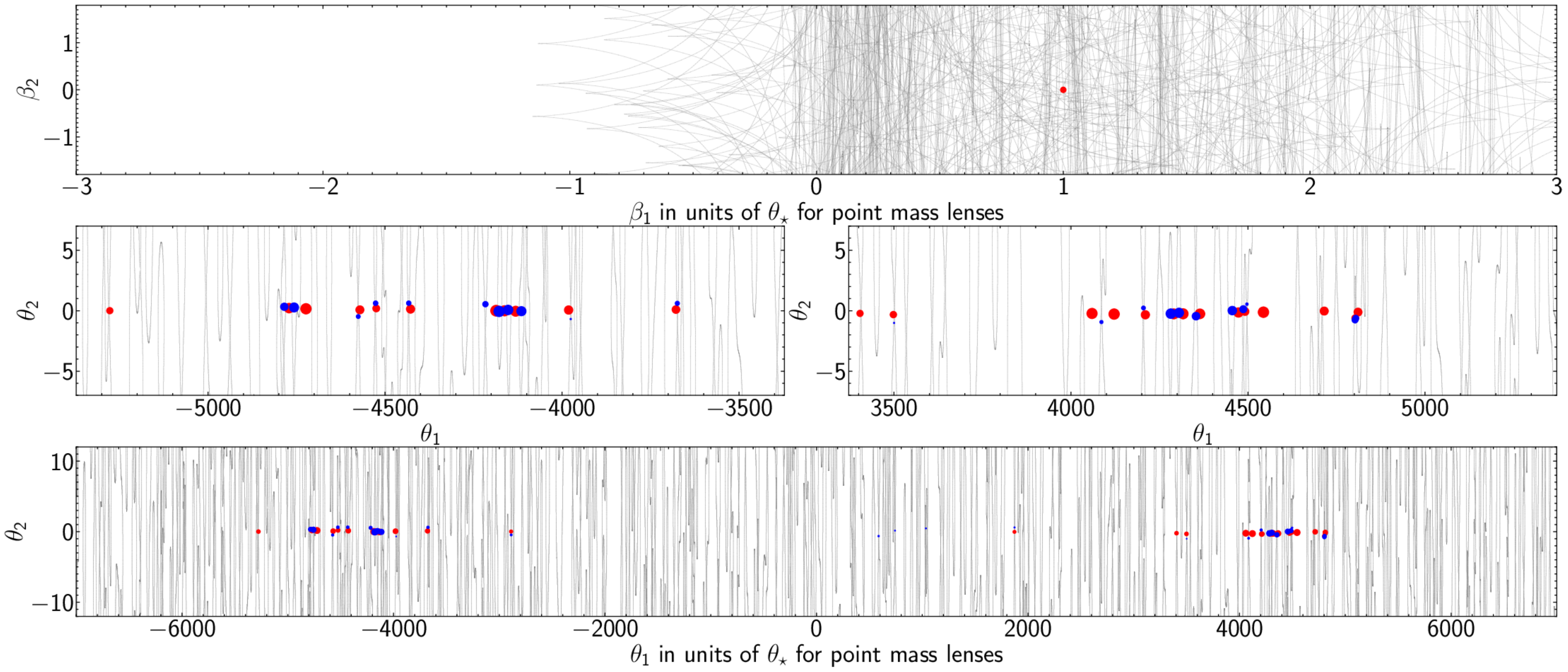}\hfill
\caption{An example of the micro-image swarms (or ``trains''). Upper and lower panels show the source and image planes, respectively, and the middle panel shows enlarged view of micro-images in the plane around the swarms. The macro model caustic and critical curve (i.e., those in absence of microlenses) are vertical lines at $\beta_1=\theta_1=0$. Micro-caustics and critical curves are shown by gray thin lines. The source position is ($\beta_1$, $\beta_2$)=($1$, $0$), and corresponding multiple images in the image plane are shown by filled circles in the middle and lower panels. The size of the circles scales with the logarithm of the magnification factor of each image, and red and blue circles indicate micro-minimum and micro-saddle point images. In total there are 12 micro-minimum images on the negative parity side ($\theta_1<0$), and 15 micro-minimum images on the positive parity side ($\theta_1>0$).
The set-up of the calculation is same as the one in Figure~\ref{fig:train_mag_all}.}
\label{fig:train_demo}
\end{figure*}

\subsection{Effective sizes}

Accounting for the high-angle tail of the PDF of the microlens deflection angle, about 99\% of the flux of a point source should be found within a region of size $10\cdot\theta_\star\kappa_\star^{1/2}$ in the source plane, when considering the smooth mapping to the source place caused only by the smoothed lens and including the average effect of microlenses as as an effective source, as explained above. Transforming to the image plane, one gets an ellipse of semi-major axes $r_1\approx 10\cdot\theta_\star\kappa_\star^{1/2}\mu_1$ and  $r_2\approx 10\cdot\theta_\star\kappa_\star^{1/2}\mu_2$. This effective size of the micro image swarm is discussed as well in \cite{2024SSRv..220...14V} for the case of quasars. 

Figure~\ref{fig:train_demo} shows an example of the micro image swarm (or ``train'') from calculations given in Section~\ref{sec:effective_extend}. The Figure indicates that there are two trains of the micro images on both sides of the macro critical curve at $\theta_1=0$ such that they are highly spread along horizontal directions. Since in our macro mass model $\mu_2=1$ and $\mu_1(\theta_1=4000)=2500$, the discussion above suggests that the size of the swarm are $r_1 \approx 1250$ and $r_2\approx 0.5$ in the dimensionless unit. The distribution of micro images shown in Figure~\ref{fig:train_demo} appears to be consistent with this estimate.

\begin{figure}
\centering
\includegraphics[width=.9\linewidth]{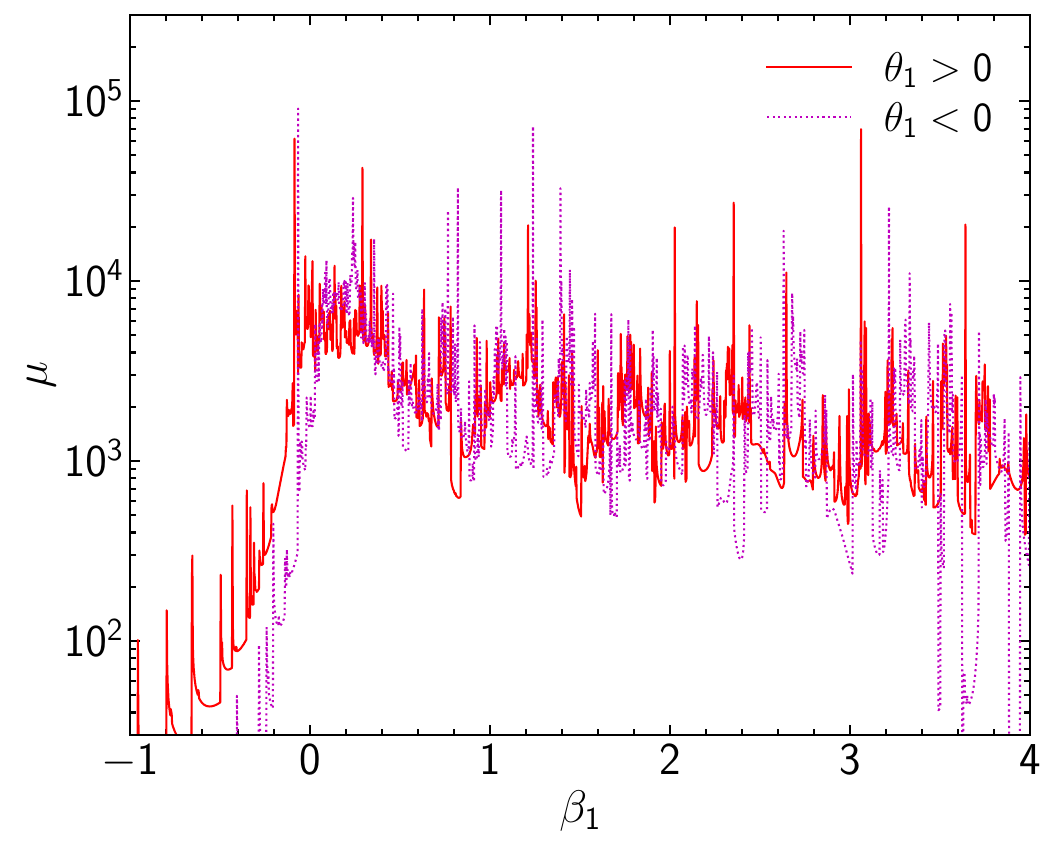}
\caption{Similar to Figure~\ref{fig:train_mag_all}, but magnification factors on the positive parity side ($\theta_1>0$) and on the negative parity side ($\theta_1<0$) are plotted separately by solid and dotted lines, respectively.}
\label{fig:train_mag_parity}
\end{figure}

\subsection{Parity dependence}

As seen in Figure~\ref{fig:train_demo}, unless $\beta_1$ is small enough that the two swarms of images are merged, there are two image swarms appearing on each side of the macro-critical line. If the two swarms are well separated, they can be observationally resolved and their total magnification can be separately measured. To check the different behavior of each swarm, Figure~\ref{fig:train_mag_parity} shows the total magnifications of the two swarms located on the positive and negative parity sides as a function of the source position. A noticeable difference is that the total magnification of the negative parity side reaches significantly smaller values. As noted in \citet{2018ApJ...857...25D} and \citet{2018PhRvD..97b3518O}, this is because in the negative parity region, a source can be more strongly de-magnified by the microlenses de-magnified compared with the original macro model magnification \citep[see also, e.g.,][]{2002ApJ...580..685S}. This effect becomes less pronounced as $\beta_1$ decreases because of averaging over many micro-images (see also Section~\ref{sec:mean_number}).

\subsection{Thickness of the corrugated macro-caustic}
\label{sec:corrugated_band}

The micro-critical lines are corrugated, that is to say, typically joined together in a large-scale network, within a width $\theta_{\rm w}$ around the original macro-critical curve, where the characteristic size of the micro-critical curve of a single microlens is equal to the mean separation between microlenses \citep{2017ApJ...850...49V}. For microlenses with Einstein radius $\theta_\star$ and convergence contribution $\kappa_\star$, their mean separation is given by $\theta_\star \kappa_\star^{-1/2}$. At a separation $\theta$ from the macro-critical curve, the magnification eigenvalue of the smooth lens is $\theta/E$, and the micro-critical curve is boosted to a size $\theta_\star (\theta/E)^{-1/2}$. Equating these two quantities we derive the thickness of the corrugated micro-critical curves, $\theta_{\rm w}$, as
\begin{equation}
     \theta_{\rm w} = E\kappa_\star~.
\end{equation}
For our specific example shown in this paper, we find $\theta_{\rm w}=25000$, which indicates that results shown in previous subsections are those for micro-images well within the corrugated band.
Within this thickness $\theta_{\rm w}$, the magnification of the individual micro-images is roughly constant, and given by roughly the magnification at the separation $\theta_{\rm w}$ i.e., 
\begin{equation}
     \mu_i \sim \frac{1}{2(1-\kappa_0)\kappa_\star}~,
     \label{eq:mean_mag_est}
\end{equation}
which reduces to $\mu_i \sim 400$ in our example. Note that this typical magnification of an individual micro-image is not the same as the total magnification, which increases as we approach the macro-critical line owing to the increasing number of micro-images in the two swarms. As we will see below, the maximum magnification achieved in microcaustic crossings is also constant within the region of width $r_w$, in the same way as the typical magnification of a single micro-image.

\begin{figure}
\centering
\includegraphics[width=.9\linewidth]{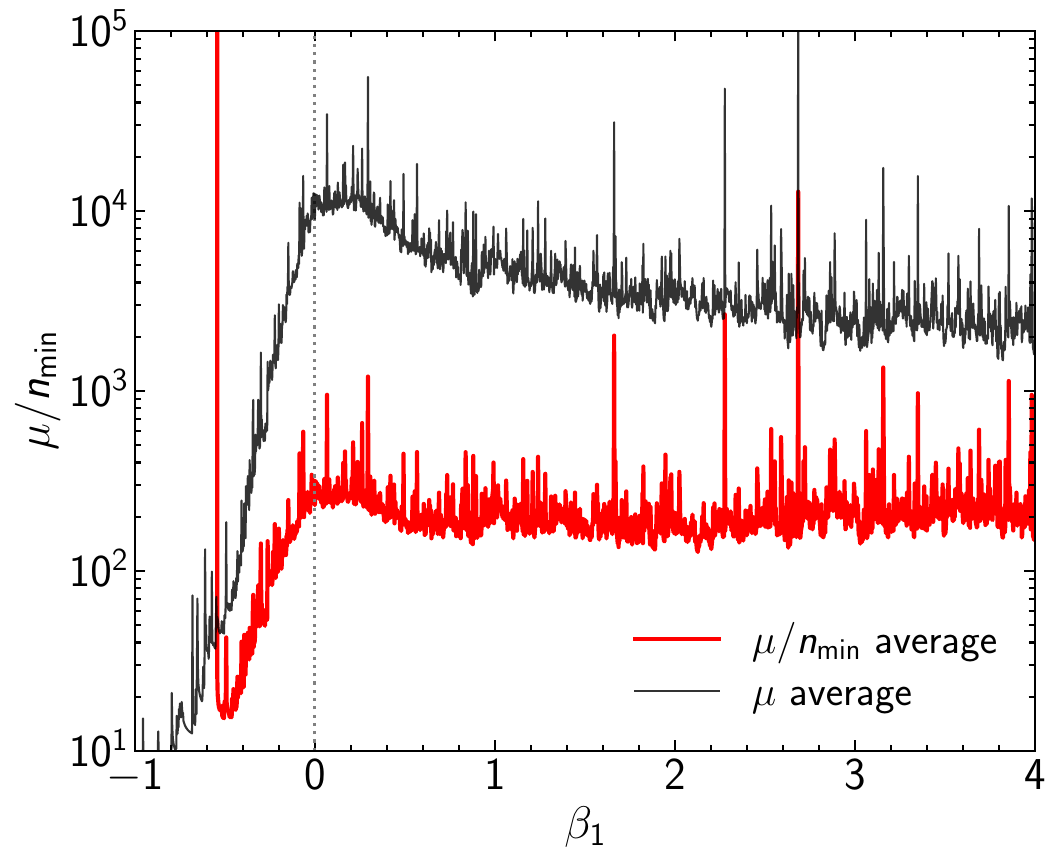}\hfill
\caption{Magnification factors divided by the number of micro-minimum images ($n_{\rm min}$), averaged over 10 realizations, are plotted as a function of the source position $\beta_1$ near the macro-caustic. For comparison, the average total magnification factors plotted in Figure~\ref{fig:train_mag_all_ave} are also shown by a thin line. The set-up of the calculation is same as the one in Figure~\ref{fig:train_mag_all}.}
\label{fig:train_mag_mave_ave}
\end{figure}

\subsection{Mean number of extra micro-image pairs}
\label{sec:mean_number}

\citet{2009JMP....50l2501P} provide a generic formula for the expected number of positive parity images in lensing systems (see their Eq.~11). In general, the formula relies on knowing the expected value of the magnification as a function of image plane position, conditional upon the fact that image plane positions are mapped to a particular source-plane position. In the case of constant surface mass density and shear, the formula simplifies greatly due to the fact that the expected value of the magnification is independent of image plane position. This is not the case near a macro-caustic, and the matter of the distribution of the magnification is currently an open question. We cannot therefore provide analytic estimates of the mean number of micro-minima, nor the mean magnification per micro-minima, as we could with constant surface mass density and shear. However, we can provide some commentary based on arguments from the simpler case. 

For constant convergence and shear that produces an image of formally infinite magnification, the equations in \cite{2024SSRv..220...14V} are found to produce an infinite number of expected micro-minima. However, the random shear due to the micro-lenses combines in such a way that the mean magnification per minimum is finite. We know that near a macro-caustic, the magnification will not in fact diverge -- as the discussion of the previous sections show, the point mass behaves as a an extended source. The number of micro-minima will not diverge then either. 

Figure~\ref{fig:train_mag_mave_ave} shows the magnification divided by the the number of micro-minima from calculations given in Section~\ref{sec:effective_extend}. The result indicates that the increase of the mean magnification as approaching the macro-caustic is achieved mainly by the increased number of micro-images, rather than the increase of magnifications of individual micro-images. It is found that the average magnification of micro-minima is $\sim 200-300$, which we find is reasonably close to a rough estimation given in equation~(\ref{eq:mean_mag_est}).

The maximum number of images is reached when the two swarms merge, over a region in the source plane with the width of the effective source size induced by microlensing, $\theta_\star \kappa_\star^{1/2}\sim 10^{-7}\, {\rm arcsec}$ for typical values. The time it would take for a source to cross this region is of the order of a year, so if a supermagnified source is close to the position of the smooth model caustic, it should be possible to measure the variation in the frequency of caustic crossings and average magnification that occurs at this position over this timescale.

\subsection{Maximum achievable magnification}\label{sec:mu_max}

%Jordi: We had discussed the main points we should aim to clarify %in this review article:
  
%  1. What is the mean magnification around a macrocaustic, and the distribution of contributions from individual images?
  
% 2. What is the difference in magnification distributions between the two asymmetric sides of the microcaustic network, and therefore between the two macroimages?
  
%  3. What are the limits on the surface density of point masses from the observed events near lensing cluster macrocaustics, two so far? Can one hope to obtain constraints also on the mass function of microlenses?
  
%  4. Can we obtain similar limits from microlensed, multiply imaged quasars? Review Mediavilla result, compare to highly magnified stars near macrocaustics.

The maximum magnification in a microimage is reached when the source crosses a micro-caustic, which occurs over a timescale determined by the source angular radius and the transverse velocities of source and lens relative to us \citep{1991ApJ...379...94M}. The maximum magnification is determined by the eigenvalue gradient $E$ at the critical line, and its characteristic value is limited by the presence of compact microlenses within the large-scale gravitational lens. The practical case this has been observed in is when the large-scale lens is a cluster of galaxies or a galaxy, and the compact microlenses are individual stars in cluster galaxies or the intracluster population.

%We will consider in general that the convergence corresponding to the surface density of microlenses is $\kappa_\star \ll 1$. We all consider the simple case when all microlenses have the same mass, and an Einstein radius $\theta_\star$ when they are isolated.

%In general, the definition of macro-critical lines for a smooth lens (in the absence of microlenses) is the set of points in the image plane where one of the two eigenvalues is zero: $1-\kappa-\gamma=0$, or $1-\kappa+\gamma=0$. In the presence of microlenses, the corrugated band is defined by the requirement $| 1-\kappa-\gamma| < \kappa_\star$ (or
%$| 1-\kappa+\gamma| < \kappa_\star$). Therefore, these corrugated bands exist around the macro-critical lines of the smooth lens. At a random point in the image plane, the nearest microlens is at a mean distance $\theta_1=\theta_\star\kappa_\star^{-1/2}$, and the mean shear that is added to that caused by the smooth macro-lens is $\delta\gamma \sim \kappa_\star$. Therefore, the micro-critical lines of individual microlenses typically merge with each other within the corrugated band, and remain separate around each microlens outside this band.

As shown in Section~\ref{sec:corrugated_band}, within the corrugated band, the average magnification of any random point (i.e., microimage) in the image plane is $\bar\mu \sim  [2\kappa_\star(1-\kappa_0)]^{-1}$.
This can also be interpreted as the eigenvalue with smallest absolute value fluctuating with a typical value of $\kappa_\star$ with the other eigenvalue being fixed to $2(1-\kappa_0)$. 
The scale of variation of the small eigenvalue in the image plane is the mean distance between microlenses, $\theta_\star\kappa_\star^{-1/2}$, which in the source plane results in a typical separation between micro-caustics of $\theta_\star\kappa_\star^{1/2}$, implying that the maximum magnification that is typically reached for individual images of a source of angular radius $\theta_{\rm s}$ in micro-critical lines is
\begin{equation}
    \mu_{\rm peak} \sim { (\theta_\star\kappa_\star^{1/2}/\theta_{\rm s})^{1/2} \over 2\kappa_\star(1-\kappa_0) } = {\theta_\star^{1/2} \kappa_\star^{-3/4} \over 2\theta_{\rm s}^{1/2}(1-\kappa_0) }~.
\end{equation}
An interesting observation from this result is that the maximum achievable magnification decreases with increasing stellar mass density $\kappa_\star$.

%\section{The impact of diffraction}

\section{Applications of observations}\label{sec:application}

\subsection{Summary of observations}\label{sec:obs}

The first discovery of a micro-caustic crossing near a macro-caustic was reported by \citet{2018NatAs...2..334K}, as the rapid transient MACS J1149 Lensed Star 1 (also known as `Icarus') near the macro critical curve of the massive cluster MACS J1149.5+2223 at $z=0.54$, interpreted as a highly magnified individual star at $z=1.49$. The timescale of the lightcurve near the peak ($\lesssim 10$~days) constrains the size of the background source to be $\lesssim 200~R_\odot$ for typical transverse velocities expected from large-scale structure \citep{2018PhRvD..97b3518O}. Together with the spectral energy distribution and peak magnitude, this led
%compatible with a blue giant suggests that the background source must be an individual star rather than e.g., a star cluster. From the size, the peak magnitude, and the spectral energy distribution, 
\citet{2018NatAs...2..334K} to conclude that the transient is an image of a blue supergiant at $z=1.49$ magnified by more than a factor of $2000$. Stars making up the intracluster light can fully account for the observed microlensing event rate. \citet{2018NatAs...2..334K} also reported a separate short transient event detected $0.26''$ from Lensed Star 1, which can be a counterimage of Lensed Star 1.

\citet{2018NatAs...2..324R} reported two peculiar fast transients (collectively nicknamed `Spock') in a giant arc at $z=1.0054$ lensed by the massive cluster MACS~J0416.1$-$2403 at $z=0.397$. While they interpret the transients as eruptions of a luminous blue variable star or a recurrent nova, they also noted these events might be micro-caustic crossings near the macro-critical curve. Improved mass modeling of this cluster to constrain the shape of the critical curves would be helpful to discriminate these possibilities.

The discovery of Icarus has triggered searches of micro-caustic crossings in archival {\it Hubble Space Telescope} images. \citet{2019ApJ...881....8C} and \citet{2019ApJ...880...58K} reported the discovery of a highly magnified star (blue supergiant) at redshift $z=0.94$ in a giant arc behind the massive cluster MACS~J0416.1$-$2403. The possible counterimage of the background star at an offset of $\sim 0.1''$ is also reported. The discovery suggests that micro-caustic crossing events may be ubiquitously found in deep imaging of massive clusters of galaxies.

\citet{2020MNRAS.495.3192D} reported several asymmetric surface brightness features in a giant arc at $z=2.93$ produced by the galaxy cluster SDSS J1226+2152 at $z=0.43$. While such asymmetric features can be produced by micro-caustic crossings, \citet{2020MNRAS.495.3192D} concluded that they are more likely to be produced by subhalos with masses of $\sim 10^6-10^8M_\odot$, based on the absence of notable time variation over a six-year baseline.

Several other observations of supermagnified stars in lensing clusters have been reported recently, some of which are being given special names, and are continuing to be discovered with the ongoing observations with JWST: an additional star called {\it Mothra} in MACSJ0416 \citep{DSY23}, the star {\it Earendel} in WHL0137-08 \citep{WCZ22,WCD22}, Godzilla in PSZ1 G311.65 \citep{DPK22}, {\it Quyllur} in the El Gordo cluster \citep{DMA23}, and several other stars in Abell 2744 \citep{CKT22}, MACSJ0647 \citep{MZJ23,Furtak24}, and Abell 370 \citep{Kelly22,MCZ23}. In this short paper we cannot do justice to all the wonderful science that is made possible by this blossoming of discoveries.

\subsection{Some applications}\label{sec:app}

We now discuss some applications that the phenomenon of super-magnification in the micro-caustics of lensing clusers of galaxies has already demonstrated or may be developed in the future. This list is by no means complete and is limited by the length of this paper; in general any source, such as supernovae or quasars, may be studied in special ways when highly magnified; in the case of gravitational waves, diffraction effects mean that the interest of lensing observations shifts to masses larger than stellar ones, corresponding to a Schwarzshild radius larger than the observed wavelengths.

\subsubsection{Compact objects in the dark matter}

As discussed in Section~\ref{sec:microlens}, one of the most important model parameters that control the property of microlensing near macro-caustics is the surface density of microlenses, $\kappa_\star$. Normally only stars in lensing clusters or galaxies are considered as microlenses, whose abundance near macro-caustics can be inferred from observations of intracluster light. However, if a fraction of dark matter is composed of dark compact objects, such as primordial black holes \citep{1974MNRAS.168..399C}, the surface density of microlenses may be much higher. This, in turn, means we can constrain the abundance of compact objects from observations of micro-caustic crossings in a manner complementary to other constraints \citep[see e.g.,][for a review of constraints on primordial black holes]{2018CQGra..35f3001S}, including quasar microlensing \citep{2017ApJ...836L..18M}.

Constraints on compact objects from the Icarus highly magnified star have extensively been explored in \citet{2018NatAs...2..334K} as well as in follow-up papers by \citet{2017ApJ...850...49V}, \citet{2018ApJ...857...25D}, and \citet{2018PhRvD..97b3518O}. For instance, as explained in Section~\ref{sec:mu_max} the maximum achievable magnification is reduced as $\kappa_\star^{-3/4}$. Following this idea, \citet{2018PhRvD..97b3518O} adopted a simple analytic model to derive constraints on the compact object abundance for a wide range of mass ranging from $10^{-5}M_\odot$ to $10^{2}M_\odot$. In addition, a large $\kappa_\star$ reduces the time variability of lensed sources due to averaging effects, which also yields constraints \citep{2018NatAs...2..334K,2018ApJ...857...25D}.

Furthermore, as discussed in Section~\ref{sec:corrugated_band}, the compact object abundance determines the width of the corrugated band around the macro-critical curve within which most highly magnified images appear, so an independent constraint is obtained from the spread of micro-caustic crossings around macro critical curves \citep{2017ApJ...850...49V}. A limit can then be derived on $\kappa_\star$ in compact objects from the set of highly magnified stars at cosmological distances reported so far.
%, as an example of the potential of this method as more of these events are discovered in the future
The width of the corrugated band is $\theta_w = d_\parallel \kappa_\star$, where $d_\parallel$ is the gradient of the small magnification eigenvalue. In the first event named Icarus, the macro model predicts $d_\parallel \simeq 4\, {\rm arcmin}^{-1}$, and the separation of the image from the predicted critical curve is $\sim 0.13$ arcsec \citep{2018NatAs...2..334K,2017ApJ...850...49V}. In the MACS J0416.1-2403 event, the macro model predicts $d_\parallel = 7\, {\rm arcmin}^{-1}$ and the separation of the image is $0.1$ arcsec \citep{2018ApJ...857...25D,2018PhRvD..97b3518O}. The ratios of the observed image separations compared to the maximum separation where images with the highest magnifications are observed is therefore $0.012\kappa_\star$ and $0.009 \kappa_\star$, respectively. The random probability these two values are observed at least as small as they are is $\sim 10^{-4}\kappa_\star^2$. Requiring this probability to be at least larger than 1\% we obtain the limit $\kappa_\star < 0.1$. This is valid for a broad range of microlens mass, roughly $10^{-6}\, M_\odot < M$, over which other effects like diffraction or source size are not important \citep[see Figure 9 in][]{2017ApJ...850...49V}. Naturally, as the number of discovered events increases in cluster lenses that can be adequately modeled, this limit on any objects that are sufficiently compact to act as microlenses near macro-caustics will improve essentially to the value of the microlenses accounted by intracluster stars.

Recently, this argument has been developed in detail using most of the discovered supermagnified stars by \cite{2024arXiv240316989V}, and in the future this limit should rapidly improve as more examples of these stars with extreme magnifications are revealed by the potential of JWST and other telescopes.

\subsubsection{Observing Population III stars}

Microlensing near the macro-caustics provides a means of observing distant individual stars that cannot be observed without the high magnification achieved in micro-caustic crossings. \citet{2018ApJS..234...41W} explored the possibility of observing individual Population III (Pop III) stars, which are metal-free stars formed from pristine gas, at $z\simeq 7-17$, and concluded that direct observations of highly magnified Pop III stars would be possible by monitoring a few to a few tens of mass clusters over a decade with the James Webb Space Telescope. 

\subsubsection{Dark matter subhalos}

\citet{2018ApJ...867...24D} proposed a novel method to constrain the abundance of dark matter subhalos with masses of $10^6-10^8M_\odot$, whose existence is naturally predicted by the standard cold dark matter paradigm \citep[e.g.,][]{2008Natur.454..735D}, using observations of image pairs of sources near caustics. 
The method uses astrometry of these highly magnified image pairs of stars or other luminous objects (star clusters or star-forming regions) in caustic-straddling giant arcs and searches for perturbations of the shape of macro critical curves, which are expected to be smooth on small scales in the absence of subhalos. Subhalos with masses of $10^6-10^8M_\odot$ produce irregularities on the critical curve shape \citep[see also][]{AKO24}, and the astrometric perturbations of image pairs, at the level of $20-80$~mas, which can be detected by accurate centroid measurements of several magnified pairs by the James Webb Space Telescope. Known cluster galaxies produce irregularities on larger scales only, corresponding to their higher masses.

\subsubsection{Axion  minihalos}

Finally, axion minihalos may also be detectable using highly magnified stars. Axion minihalos are expected in many variations of the theory of axions for dark matter that solve the strong QCD problem, with masses of $\sim 10^{-12}\, M_\odot$. These are extremely difficult to detect, not only because of their tiny size but also because their surface density is highly subcritical to lensing. However, when a star crosses one of the micro-caustics affected by micro-lensing from intracluster stars, axion minihalos would create a secondary level of perturbations on the magnification at smaller scales that would cause perturbations of the lightcurve of micro-caustic crossing events on small time scales. This was predicted in \citet{2020AJ....159...49D}, proposing a possible route to the detection of axion dark matter using only astrophysical methods.

\section{Summary}\label{sec:summary}

In this chapter we have explored the expected microlensing magnification of sources close to macro-caustics. For this purpose, we have performed simulations and compared the results to the analytical expectations of the behavior of microimages, also described in \cite{2017ApJ...850...49V, 2021arXiv210412009D, 2024SSRv..220...14V}. 

We have shown that the mean magnification due to lens plane microlenses of point sources in the vicinity of the macro-caustic follows that of an extended source in the vicinity of a macro-caustic in the absence of microlensing. This is precisely correct when the source profile matches the probability distribution function of the microlens deflection angle. 

Similarly, we have shown that the two swarms of microimages at each side of the macro-caustic are spread within an ellipse that results from mapping the microlens deflection angle PDF, or the corresponding ``extended source'', to the lens plane. We note that even though the sizes of the two image swarms are symmetric, the swarm of images at the negative parity side of the macro-caustic can show significantly lower magnification values. This is consistent with the findings of \cite{2002ApJ...580..685S} who show that microlensed sources in macro-saddle points are more prone to be de-magnified than in macro-minima.

We have reviewed the characteristic number and average magnification of the micro-images within the microlens-distorted (or corrugated) macro-caustic. We observe that the magnification per micro-image remains roughly constant within this corrugated caustic and with a value given by the separation of micro-image pairs from micro-caustics. The maximum individual micro-image magnification is limited by the angular size of the source, and decreases with increasing stellar mass density. The increase of the mean magnification when approaching the position of the macro-caustic is driven by the increase in the number of micro-images rather than their individual magnifications. Analogous to the fact that micro-images behave as a lensed extended source and thus the total magnification is not infinite when crossing the macro-caustic, the number of micro-images reaches a maximum when the two image swarms merge.

The microlensing phenomena occurring near macro-caustics that we have reviewed in this Chapter have not been observed until recently. Since the dawn of the era of time-domain astronomy, in the past few years several transients have been shown to be consistent with such microlensing in the vicinity of macro-caustics. These discoveries imply that microcaustic crossing events should be ubiquitous when monitoring observations of massive galaxy clusters, and will surely be increased by observations with JWST reaching for fainter magnitudes. With this in mind, we have non-exhaustively reviewed some future (and current) applications of the different observational signatures, ranging from identifying population III stars from individual caustic crossing events to constraining the abundance and nature of primordial black holes and dark matter subhalos from the statistical properties of observed caustic crossing events and probing the distorted macro-caustic from astrometric perturbations in magnified background stars.

\begin{acknowledgements}
This work was supported in part by JSPS KAKENHI Grant Number JP22H01260, JP20H05856, and by Spanish grant PID2019-108122GB-C32. TA acknowledges support the Millennium Science Initiative ICN12\_009 and the ANID BASAL project FB210003.
\end{acknowledgements}

\section*{Conflict of Interest Statement}
The authors declare no competing interests.

%\begin{figure}
%\centering
%\includegraphics[width=.3\linewidth]{figs/macromin.png}
%\includegraphics[width=.3\linewidth]{figs/smalltrain.png}
%\includegraphics[width=.3\linewidth]{figs/macromax.png}
%\caption{Smalls train of micro-images near (not too near) macro-caustics. Left: on the macro-minimum side, middle: on the macro-saddle side, right: on the macro-maximum side.}
%\label{fig:smalltrain}
%\end{figure}

\def\aj{AJ}
\def\apj{APJ}
\def\apjs{APJS}
\def\mnras{MNRAS}
\def\prd{PRD}

\bibliographystyle{aps-nameyear}
\bibliography{references}

\begin{thebibliography}{41}
% BibTex style file: aps.bst  (nameyear), 2017-12-15
\ifx \bisbn   \undefined \def \bisbn  #1{ISBN #1}\fi
\ifx \binits  \undefined \def \binits#1{#1} \fi
\ifx \bauthor  \undefined \def \bauthor#1{#1} \fi
\ifx \bjtitle  \undefined \def \bjtitle#1{\textrm{#1}}\fi
\ifx \batitle  \undefined \def \batitle#1{#1} \fi
\ifx \bctitle  \undefined \def \bctitle#1{#1} \fi
\ifx \bvolume  \undefined \def \bvolume#1{\textbf{#1}}\fi
\ifx \byear  \undefined \def \byear#1{#1} \fi
\ifx \bissue  \undefined \def \bissue#1{#1} \fi
\ifx \bfpage  \undefined \def \bfpage#1{#1} \fi
\ifx \blpage  \undefined \def \blpage #1{#1} \fi
\ifx \burl  \undefined \def \burl#1{#1} \fi
\ifx \doiurl  \undefined \def \doiurl#1{#1} \fi
\ifx \betal  \undefined \def \betal{et al.} \fi
\ifx \binstitute  \undefined \def \binstitute#1{#1} \fi
\ifx \beditor  \undefined \def \beditor#1{#1} \fi
\ifx \bpublisher  \undefined \def \bpublisher#1{#1} \fi
\ifx \bbtitle  \undefined \def \bbtitle#1{\textit{#1}} \fi
\ifx \bedition  \undefined \def \bedition#1{#1} \fi
\ifx \bseriesno  \undefined \def \bseriesno#1{#1} \fi
\ifx \blocation  \undefined \def \blocation#1{#1} \fi
\ifx \bsertitle  \undefined \def \bsertitle#1{#1} \fi
\ifx \bsnm \undefined \def \bsnm#1{#1} \fi
\ifx \bsuffix \undefined \def \bsuffix#1{#1} \fi
\ifx \bparticle \undefined \def \bparticle#1{#1} \fi
\ifx \barticle \undefined \def \barticle#1{#1} \fi
\ifx \botherref \undefined \def \botherref #1{#1} \fi
\ifx \url \undefined \def \url#1{#1} \fi
\ifx \bchapter \undefined \def \bchapter#1{#1} \fi
\ifx \bbook \undefined \def \bbook#1{#1} \fi
\ifx \bcomment \undefined \def \bcomment#1{#1} \fi
\ifx \oauthor \undefined \def \oauthor#1{#1} \fi
\ifx \citeauthoryear \undefined \def \citeauthoryear#1{#1} \fi
\ifx \texttildelow  \undefined \def \texttildelow{\symbol{126}} \fi
\def \endbibitem {}
\ifx \bconflocation  \undefined \def \bconflocation#1{#1} \fi

\bibitem[\protect\citeauthoryear{{Abe} et~al.}{2023}]{AKO24}
\begin{botherref}
\oauthor{\binits{K.T.} \bsnm{{Abe}}},
\oauthor{\binits{H.} \bsnm{{Kawai}}},
\oauthor{\binits{M.} \bsnm{{Oguri}}},
{Analytic approach to astrometric perturbations of critical curves by
  substructures}.
arXiv e-prints,
2311--18211
(2023).
\doiurl{https://doi.org/10.48550/arXiv.2311.18211}
\end{botherref}
\endbibitem

\bibitem[\protect\citeauthoryear{{Carr} and
  {Hawking}}{1974}]{1974MNRAS.168..399C}
\begin{barticle}
\bauthor{\binits{B.J.} \bsnm{{Carr}}},
\bauthor{\binits{S.W.} \bsnm{{Hawking}}},
\batitle{{Black holes in the early Universe}}.
\bjtitle{\mnras}
\bvolume{168},
\bfpage{399}--\blpage{416}
(\byear{1974}).
\doiurl{https://doi.org/10.1093/mnras/168.2.399}
\end{barticle}
\endbibitem

\bibitem[\protect\citeauthoryear{{Chen} et~al.}{2019}]{2019ApJ...881....8C}
\begin{barticle}
\bauthor{\binits{W.} \bsnm{{Chen}}},
\bauthor{\binits{P.L.} \bsnm{{Kelly}}},
\bauthor{\binits{J.M.} \bsnm{{Diego}}},
\bauthor{\binits{M.} \bsnm{{Oguri}}},
\bauthor{\binits{L.L.R.} \bsnm{{Williams}}},
\bauthor{\binits{A.} \bsnm{{Zitrin}}},
\bauthor{\binits{T.L.} \bsnm{{Treu}}},
\bauthor{\binits{N.} \bsnm{{Smith}}},
\bauthor{\binits{T.J.} \bsnm{{Broadhurst}}},
\bauthor{\binits{N.} \bsnm{{Kaiser}}},
\bauthor{\binits{R.J.} \bsnm{{Foley}}},
\bauthor{\binits{A.V.} \bsnm{{Filippenko}}},
\bauthor{\binits{L.} \bsnm{{Salo}}},
\bauthor{\binits{J.} \bsnm{{Hjorth}}},
\bauthor{\binits{J.} \bsnm{{Selsing}}},
\batitle{{Searching for Highly Magnified Stars at Cosmological Distances:
  Discovery of a Redshift 0.94 Blue Supergiant in Archival Images of the Galaxy
  Cluster MACS J0416.1-2403}}.
\bjtitle{\apj}
\bvolume{881}(\bissue{1}),
\bfpage{8}
(\byear{2019}).
\doiurl{https://doi.org/10.3847/1538-4357/ab297d}
\end{barticle}
\endbibitem

\bibitem[\protect\citeauthoryear{{Chen} et~al.}{2022}]{CKT22}
\begin{barticle}
\bauthor{\binits{W.} \bsnm{{Chen}}},
\bauthor{\binits{P.L.} \bsnm{{Kelly}}},
\bauthor{\binits{T.} \bsnm{{Treu}}},
\bauthor{\binits{X.} \bsnm{{Wang}}},
\bauthor{\binits{G.} \bsnm{{Roberts-Borsani}}},
\bauthor{\binits{A.} \bsnm{{Keen}}},
\bauthor{\binits{R.A.} \bsnm{{Windhorst}}},
\bauthor{\binits{R.} \bsnm{{Zhou}}},
\bauthor{\binits{M.} \bsnm{{Bradac}}},
\bauthor{\binits{G.} \bsnm{{Brammer}}},
\bauthor{\binits{V.} \bsnm{{Strait}}},
\bauthor{\binits{T.J.} \bsnm{{Broadhurst}}},
\bauthor{\binits{J.M.} \bsnm{{Diego}}},
\bauthor{\binits{B.L.} \bsnm{{Frye}}},
\bauthor{\binits{A.K.} \bsnm{{Meena}}},
\bauthor{\binits{A.} \bsnm{{Zitrin}}},
\bauthor{\binits{M.} \bsnm{{Pascale}}},
\bauthor{\binits{M.} \bsnm{{Castellano}}},
\bauthor{\binits{D.} \bsnm{{Marchesini}}},
\bauthor{\binits{T.} \bsnm{{Morishita}}},
\bauthor{\binits{L.} \bsnm{{Yang}}},
\batitle{{Early Results from GLASS-JWST. VIII. An Extremely Magnified Blue
  Supergiant Star at Redshift 2.65 in the A2744 Cluster Field}}.
\bjtitle{\apjl}
\bvolume{940}(\bissue{2}),
\bfpage{54}
(\byear{2022}).
\doiurl{https://doi.org/10.3847/2041-8213/ac9585}
\end{barticle}
\endbibitem

\bibitem[\protect\citeauthoryear{{Dai} and
  {Miralda-Escud{\'e}}}{2020}]{2020AJ....159...49D}
\begin{barticle}
\bauthor{\binits{L.} \bsnm{{Dai}}},
\bauthor{\binits{J.} \bsnm{{Miralda-Escud{\'e}}}},
\batitle{{Gravitational Lensing Signatures of Axion Dark Matter Minihalos in
  Highly Magnified Stars}}.
\bjtitle{\aj}
\bvolume{159}(\bissue{2}),
\bfpage{49}
(\byear{2020}).
\doiurl{https://doi.org/10.3847/1538-3881/ab5e83}
\end{barticle}
\endbibitem

\bibitem[\protect\citeauthoryear{{Dai} and
  {Pascale}}{2021}]{2021arXiv210412009D}
\begin{botherref}
\oauthor{\binits{L.} \bsnm{{Dai}}},
\oauthor{\binits{M.} \bsnm{{Pascale}}},
{New Approximation of Magnification Statistics for Random Microlensing of
  Magnified Sources}.
arXiv e-prints,
2104--12009
(2021)
\end{botherref}
\endbibitem

\bibitem[\protect\citeauthoryear{{Dai} et~al.}{2018}]{2018ApJ...867...24D}
\begin{barticle}
\bauthor{\binits{L.} \bsnm{{Dai}}},
\bauthor{\binits{T.} \bsnm{{Venumadhav}}},
\bauthor{\binits{A.A.} \bsnm{{Kaurov}}},
\bauthor{\binits{J.} \bsnm{{Miralda-Escud}}},
\batitle{{Probing Dark Matter Subhalos in Galaxy Clusters Using Highly
  Magnified Stars}}.
\bjtitle{\apj}
\bvolume{867}(\bissue{1}),
\bfpage{24}
(\byear{2018}).
\doiurl{https://doi.org/10.3847/1538-4357/aae478}
\end{barticle}
\endbibitem

\bibitem[\protect\citeauthoryear{{Dai} et~al.}{2020}]{2020MNRAS.495.3192D}
\begin{barticle}
\bauthor{\binits{L.} \bsnm{{Dai}}},
\bauthor{\binits{A.A.} \bsnm{{Kaurov}}},
\bauthor{\binits{K.} \bsnm{{Sharon}}},
\bauthor{\binits{M.} \bsnm{{Florian}}},
\bauthor{\binits{J.} \bsnm{{Miralda-Escud{\'e}}}},
\bauthor{\binits{T.} \bsnm{{Venumadhav}}},
\bauthor{\binits{B.} \bsnm{{Frye}}},
\bauthor{\binits{J.R.} \bsnm{{Rigby}}},
\bauthor{\binits{M.} \bsnm{{Bayliss}}},
\batitle{{Asymmetric surface brightness structure of caustic crossing arc in
  SDSS J1226+2152: a case for dark matter substructure}}.
\bjtitle{\mnras}
\bvolume{495}(\bissue{3}),
\bfpage{3192}--\blpage{3208}
(\byear{2020}).
\doiurl{https://doi.org/10.1093/mnras/staa1355}
\end{barticle}
\endbibitem

\bibitem[\protect\citeauthoryear{{Diego} et~al.}{2022}]{DPK22}
\begin{barticle}
\bauthor{\binits{J.M.} \bsnm{{Diego}}},
\bauthor{\binits{M.} \bsnm{{Pascale}}},
\bauthor{\binits{B.J.} \bsnm{{Kavanagh}}},
\bauthor{\binits{P.} \bsnm{{Kelly}}},
\bauthor{\binits{L.} \bsnm{{Dai}}},
\bauthor{\binits{B.} \bsnm{{Frye}}},
\bauthor{\binits{T.} \bsnm{{Broadhurst}}},
\batitle{{Godzilla, a monster lurks in the Sunburst galaxy}}.
\bjtitle{\aap}
\bvolume{665},
\bfpage{134}
(\byear{2022}).
\doiurl{https://doi.org/10.1051/0004-6361/202243605}
\end{barticle}
\endbibitem

\bibitem[\protect\citeauthoryear{{Diego} et~al.}{2023a}]{DMA23}
\begin{barticle}
\bauthor{\binits{J.M.} \bsnm{{Diego}}},
\bauthor{\binits{A.K.} \bsnm{{Meena}}},
\bauthor{\binits{N.J.} \bsnm{{Adams}}},
\bauthor{\binits{T.} \bsnm{{Broadhurst}}},
\bauthor{\binits{L.} \bsnm{{Dai}}},
\bauthor{\binits{D.} \bsnm{{Coe}}},
\bauthor{\binits{B.} \bsnm{{Frye}}},
\bauthor{\binits{P.} \bsnm{{Kelly}}},
\bauthor{\binits{A.M.} \bsnm{{Koekemoer}}},
\bauthor{\binits{M.} \bsnm{{Pascale}}},
\bauthor{\binits{S.P.} \bsnm{{Willner}}},
\bauthor{\binits{E.} \bsnm{{Zackrisson}}},
\bauthor{\binits{A.} \bsnm{{Zitrin}}},
\bauthor{\binits{R.A.} \bsnm{{Windhorst}}},
\bauthor{\binits{S.H.} \bsnm{{Cohen}}},
\bauthor{\binits{R.A.} \bsnm{{Jansen}}},
\bauthor{\binits{J.} \bsnm{{Summers}}},
\bauthor{\binits{S.} \bsnm{{Tompkins}}},
\bauthor{\binits{C.J.} \bsnm{{Conselice}}},
\bauthor{\binits{S.P.} \bsnm{{Driver}}},
\bauthor{\binits{H.} \bsnm{{Yan}}},
\bauthor{\binits{N.} \bsnm{{Grogin}}},
\bauthor{\binits{M.A.} \bsnm{{Marshall}}},
\bauthor{\binits{N.} \bsnm{{Pirzkal}}},
\bauthor{\binits{A.} \bsnm{{Robotham}}},
\bauthor{\binits{R.E.} \bsnm{{Ryan}}},
\bauthor{\binits{C.N.A.} \bsnm{{Willmer}}},
\bauthor{\binits{L.D.} \bsnm{{Bradley}}},
\bauthor{\binits{G.} \bsnm{{Caminha}}},
\bauthor{\binits{K.} \bsnm{{Caputi}}},
\bauthor{\binits{T.} \bsnm{{Carleton}}},
\bauthor{\binits{P.} \bsnm{{Kamieneski}}},
\batitle{{JWST's PEARLS: A new lens model for ACT-CL J0102{\ensuremath{-}}4915,
  ``El Gordo,'' and the first red supergiant star at cosmological distances
  discovered by JWST}}.
\bjtitle{\aap}
\bvolume{672},
\bfpage{3}
(\byear{2023}a).
\doiurl{https://doi.org/10.1051/0004-6361/202245238}
\end{barticle}
\endbibitem

\bibitem[\protect\citeauthoryear{{Diego} et~al.}{2023b}]{DSY23}
\begin{botherref}
\oauthor{\binits{J.M.} \bsnm{{Diego}}},
\oauthor{\binits{B.} \bsnm{{Sun}}},
\oauthor{\binits{H.} \bsnm{{Yan}}},
\oauthor{\binits{L.J.} \bsnm{{Furtak}}},
\oauthor{\binits{E.} \bsnm{{Zackrisson}}},
\oauthor{\binits{L.} \bsnm{{Dai}}},
\oauthor{\binits{P.} \bsnm{{Kelly}}},
\oauthor{\binits{M.} \bsnm{{Nonino}}},
\oauthor{\binits{N.} \bsnm{{Adams}}},
\oauthor{\binits{A.K.} \bsnm{{Meena}}},
\oauthor{\binits{S.P.} \bsnm{{Willner}}},
\oauthor{\binits{A.} \bsnm{{Zitrin}}},
\oauthor{\binits{S.H.} \bsnm{{Cohen}}},
\oauthor{\binits{J.C.J.D.} \bsnm{{Silva}}},
\oauthor{\binits{R.A.} \bsnm{{Jansen}}},
\oauthor{\binits{J.} \bsnm{{Summers}}},
\oauthor{\binits{R.A.} \bsnm{{Windhorst}}},
\oauthor{\binits{D.} \bsnm{{Coe}}},
\oauthor{\binits{C.J.} \bsnm{{Conselice}}},
\oauthor{\binits{S.P.} \bsnm{{Driver}}},
\oauthor{\binits{B.} \bsnm{{Frye}}},
\oauthor{\binits{N.A.} \bsnm{{Grogin}}},
\oauthor{\binits{A.M.} \bsnm{{Koekemoer}}},
\oauthor{\binits{M.A.} \bsnm{{Marshall}}},
\oauthor{\binits{N.} \bsnm{{Pirzkal}}},
\oauthor{\binits{A.} \bsnm{{Robotham}}},
\oauthor{\binits{M.J.} \bsnm{{Rutkowski}}},
\oauthor{\binits{J.} \bsnm{{Ryan}} \bsuffix{Russell~E.}},
\oauthor{\binits{S.} \bsnm{{Tompkins}}},
\oauthor{\binits{C.N.A.} \bsnm{{Willmer}}},
\oauthor{\binits{R.} \bsnm{{Bhatawdekar}}},
{JWST's PEARLS: Mothra, a new kaiju star at z=2.091 extremely magnified by
  MACS0416, and implications for dark matter models}.
arXiv e-prints,
2307--10363
(2023b).
\doiurl{https://doi.org/10.48550/arXiv.2307.10363}
\end{botherref}
\endbibitem

\bibitem[\protect\citeauthoryear{{Diego} et~al.}{2018}]{2018ApJ...857...25D}
\begin{barticle}
\bauthor{\binits{J.M.} \bsnm{{Diego}}},
\bauthor{\binits{N.} \bsnm{{Kaiser}}},
\bauthor{\binits{T.} \bsnm{{Broadhurst}}},
\bauthor{\binits{P.L.} \bsnm{{Kelly}}},
\bauthor{\binits{S.} \bsnm{{Rodney}}},
\bauthor{\binits{T.} \bsnm{{Morishita}}},
\bauthor{\binits{M.} \bsnm{{Oguri}}},
\bauthor{\binits{T.W.} \bsnm{{Ross}}},
\bauthor{\binits{A.} \bsnm{{Zitrin}}},
\bauthor{\binits{M.} \bsnm{{Jauzac}}},
\bauthor{\binits{J.} \bsnm{{Richard}}},
\bauthor{\binits{L.} \bsnm{{Williams}}},
\bauthor{\binits{J.} \bsnm{{Vega-Ferrero}}},
\bauthor{\binits{B.} \bsnm{{Frye}}},
\bauthor{\binits{A.V.} \bsnm{{Filippenko}}},
\batitle{{Dark Matter under the Microscope: Constraining Compact Dark Matter
  with Caustic Crossing Events}}.
\bjtitle{\apj}
\bvolume{857}(\bissue{1}),
\bfpage{25}
(\byear{2018}).
\doiurl{https://doi.org/10.3847/1538-4357/aab617}
\end{barticle}
\endbibitem

\bibitem[\protect\citeauthoryear{{Diemand} et~al.}{2008}]{2008Natur.454..735D}
\begin{barticle}
\bauthor{\binits{J.} \bsnm{{Diemand}}},
\bauthor{\binits{M.} \bsnm{{Kuhlen}}},
\bauthor{\binits{P.} \bsnm{{Madau}}},
\bauthor{\binits{M.} \bsnm{{Zemp}}},
\bauthor{\binits{B.} \bsnm{{Moore}}},
\bauthor{\binits{D.} \bsnm{{Potter}}},
\bauthor{\binits{J.} \bsnm{{Stadel}}},
\batitle{{Clumps and streams in the local dark matter distribution}}.
\bjtitle{\nat}
\bvolume{454}(\bissue{7205}),
\bfpage{735}--\blpage{738}
(\byear{2008}).
\doiurl{https://doi.org/10.1038/nature07153}
\end{barticle}
\endbibitem

\bibitem[\protect\citeauthoryear{{Furtak} et~al.}{2024}]{Furtak24}
\begin{barticle}
\bauthor{\binits{L.J.} \bsnm{{Furtak}}},
\bauthor{\binits{A.K.} \bsnm{{Meena}}},
\bauthor{\binits{E.} \bsnm{{Zackrisson}}},
\bauthor{\binits{A.} \bsnm{{Zitrin}}},
\bauthor{\binits{G.B.} \bsnm{{Brammer}}},
\bauthor{\binits{D.} \bsnm{{Coe}}},
\bauthor{\binits{J.M.} \bsnm{{Diego}}},
\bauthor{\binits{J.J.} \bsnm{{Eldridge}}},
\bauthor{\binits{Y.} \bsnm{{Jim{\'e}nez-Teja}}},
\bauthor{\binits{V.} \bsnm{{Kokorev}}},
\bauthor{\binits{M.} \bsnm{{Ricotti}}},
\bauthor{\binits{B.} \bsnm{{Welch}}},
\bauthor{\binits{R.A.} \bsnm{{Windhorst}}},
\bauthor{\bsnm{{Abdurro'uf}}},
\bauthor{\binits{F.} \bsnm{{Andrade-Santos}}},
\bauthor{\binits{R.} \bsnm{{Bhatawdekar}}},
\bauthor{\binits{L.D.} \bsnm{{Bradley}}},
\bauthor{\binits{T.} \bsnm{{Broadhurst}}},
\bauthor{\binits{W.} \bsnm{{Chen}}},
\bauthor{\binits{C.J.} \bsnm{{Conselice}}},
\bauthor{\binits{P.} \bsnm{{Dayal}}},
\bauthor{\binits{B.L.} \bsnm{{Frye}}},
\bauthor{\binits{S.} \bsnm{{Fujimoto}}},
\bauthor{\binits{T.Y.-Y.} \bsnm{{Hsiao}}},
\bauthor{\binits{P.L.} \bsnm{{Kelly}}},
\bauthor{\binits{G.} \bsnm{{Mahler}}},
\bauthor{\binits{N.} \bsnm{{Mandelker}}},
\bauthor{\binits{C.} \bsnm{{Norman}}},
\bauthor{\binits{M.} \bsnm{{Oguri}}},
\bauthor{\binits{N.} \bsnm{{Pirzkal}}},
\bauthor{\binits{M.} \bsnm{{Postman}}},
\bauthor{\binits{S.} \bsnm{{Ravindranath}}},
\bauthor{\binits{E.} \bsnm{{Vanzella}}},
\bauthor{\binits{S.M.} \bsnm{{Wilkins}}},
\batitle{{Reaching for the stars - JWST/NIRSpec spectroscopy of a lensed star
  candidate at z = 4.76}}.
\bjtitle{\mnras}
\bvolume{527}(\bissue{1}),
\bfpage{7}--\blpage{13}
(\byear{2024}).
\doiurl{https://doi.org/10.1093/mnrasl/slad135}
\end{barticle}
\endbibitem

\bibitem[\protect\citeauthoryear{{Granot} et~al.}{2003}]{2003ApJ...583..575G}
\begin{barticle}
\bauthor{\binits{J.} \bsnm{{Granot}}},
\bauthor{\binits{P.L.} \bsnm{{Schechter}}},
\bauthor{\binits{J.} \bsnm{{Wambsganss}}},
\batitle{{The Mean Number of Extra Microimage Pairs for Macrolensed Quasars}}.
\bjtitle{\apj}
\bvolume{583}(\bissue{2}),
\bfpage{575}--\blpage{583}
(\byear{2003}).
\doiurl{https://doi.org/10.1086/345447}
\end{barticle}
\endbibitem

\bibitem[\protect\citeauthoryear{{Katz} et~al.}{1986}]{1986ApJ...306....2K}
\begin{barticle}
\bauthor{\binits{N.} \bsnm{{Katz}}},
\bauthor{\binits{S.} \bsnm{{Balbus}}},
\bauthor{\binits{B.} \bsnm{{Paczynski}}},
\batitle{{Random Scattering Approach to Gravitational Microlensing}}.
\bjtitle{\apj}
\bvolume{306},
\bfpage{2}
(\byear{1986}).
\doiurl{https://doi.org/10.1086/164313}
\end{barticle}
\endbibitem

\bibitem[\protect\citeauthoryear{{Kaurov} et~al.}{2019}]{2019ApJ...880...58K}
\begin{barticle}
\bauthor{\binits{A.A.} \bsnm{{Kaurov}}},
\bauthor{\binits{L.} \bsnm{{Dai}}},
\bauthor{\binits{T.} \bsnm{{Venumadhav}}},
\bauthor{\binits{J.} \bsnm{{Miralda-Escud{\'e}}}},
\bauthor{\binits{B.} \bsnm{{Frye}}},
\batitle{{Highly Magnified Stars in Lensing Clusters: New Evidence in a Galaxy
  Lensed by MACS J0416.1-2403}}.
\bjtitle{\apj}
\bvolume{880}(\bissue{1}),
\bfpage{58}
(\byear{2019}).
\doiurl{https://doi.org/10.3847/1538-4357/ab2888}
\end{barticle}
\endbibitem

\bibitem[\protect\citeauthoryear{{Kelly} et~al.}{2018}]{2018NatAs...2..334K}
\begin{barticle}
\bauthor{\binits{P.L.} \bsnm{{Kelly}}},
\bauthor{\binits{J.M.} \bsnm{{Diego}}},
\bauthor{\binits{S.} \bsnm{{Rodney}}},
\bauthor{\binits{N.} \bsnm{{Kaiser}}},
\bauthor{\binits{T.} \bsnm{{Broadhurst}}},
\bauthor{\binits{A.} \bsnm{{Zitrin}}},
\bauthor{\binits{T.} \bsnm{{Treu}}},
\bauthor{\binits{P.G.} \bsnm{{P{\'e}rez-Gonz{\'a}lez}}},
\bauthor{\binits{T.} \bsnm{{Morishita}}},
\bauthor{\binits{M.} \bsnm{{Jauzac}}},
\bauthor{\binits{J.} \bsnm{{Selsing}}},
\bauthor{\binits{M.} \bsnm{{Oguri}}},
\bauthor{\binits{L.} \bsnm{{Pueyo}}},
\bauthor{\binits{T.W.} \bsnm{{Ross}}},
\bauthor{\binits{A.V.} \bsnm{{Filippenko}}},
\bauthor{\binits{N.} \bsnm{{Smith}}},
\bauthor{\binits{J.} \bsnm{{Hjorth}}},
\bauthor{\binits{S.B.} \bsnm{{Cenko}}},
\bauthor{\binits{X.} \bsnm{{Wang}}},
\bauthor{\binits{D.A.} \bsnm{{Howell}}},
\bauthor{\binits{J.} \bsnm{{Richard}}},
\bauthor{\binits{B.L.} \bsnm{{Frye}}},
\bauthor{\binits{S.W.} \bsnm{{Jha}}},
\bauthor{\binits{R.J.} \bsnm{{Foley}}},
\bauthor{\binits{C.} \bsnm{{Norman}}},
\bauthor{\binits{M.} \bsnm{{Bradac}}},
\bauthor{\binits{W.} \bsnm{{Zheng}}},
\bauthor{\binits{G.} \bsnm{{Brammer}}},
\bauthor{\binits{A.M.} \bsnm{{Benito}}},
\bauthor{\binits{A.} \bsnm{{Cava}}},
\bauthor{\binits{L.} \bsnm{{Christensen}}},
\bauthor{\binits{S.E.} \bsnm{{de Mink}}},
\bauthor{\binits{O.} \bsnm{{Graur}}},
\bauthor{\binits{C.} \bsnm{{Grillo}}},
\bauthor{\binits{R.} \bsnm{{Kawamata}}},
\bauthor{\binits{J.-P.} \bsnm{{Kneib}}},
\bauthor{\binits{T.} \bsnm{{Matheson}}},
\bauthor{\binits{C.} \bsnm{{McCully}}},
\bauthor{\binits{M.} \bsnm{{Nonino}}},
\bauthor{\binits{I.} \bsnm{{P{\'e}rez-Fournon}}},
\bauthor{\binits{A.G.} \bsnm{{Riess}}},
\bauthor{\binits{P.} \bsnm{{Rosati}}},
\bauthor{\binits{K.B.} \bsnm{{Schmidt}}},
\bauthor{\binits{K.} \bsnm{{Sharon}}},
\bauthor{\binits{B.J.} \bsnm{{Weiner}}},
\batitle{{Extreme magnification of an individual star at redshift 1.5 by a
  galaxy-cluster lens}}.
\bjtitle{Nature Astronomy}
\bvolume{2},
\bfpage{334}--\blpage{342}
(\byear{2018}).
\doiurl{https://doi.org/10.1038/s41550-018-0430-3}
\end{barticle}
\endbibitem

\bibitem[\protect\citeauthoryear{{Kelly} et~al.}{2022}]{Kelly22}
\begin{botherref}
\oauthor{\binits{P.L.} \bsnm{{Kelly}}},
\oauthor{\binits{W.} \bsnm{{Chen}}},
\oauthor{\binits{A.} \bsnm{{Alfred}}},
\oauthor{\binits{T.J.} \bsnm{{Broadhurst}}},
\oauthor{\binits{J.M.} \bsnm{{Diego}}},
\oauthor{\binits{N.} \bsnm{{Emami}}},
\oauthor{\binits{A.V.} \bsnm{{Filippenko}}},
\oauthor{\binits{A.} \bsnm{{Keen}}},
\oauthor{\binits{S.} \bsnm{{Kei Li}}},
\oauthor{\binits{J.} \bsnm{{Lim}}},
\oauthor{\binits{A.K.} \bsnm{{Meena}}},
\oauthor{\binits{M.} \bsnm{{Oguri}}},
\oauthor{\binits{C.} \bsnm{{Scarlata}}},
\oauthor{\binits{T.} \bsnm{{Treu}}},
\oauthor{\binits{H.} \bsnm{{Williams}}},
\oauthor{\binits{L.L.R.} \bsnm{{Williams}}},
\oauthor{\binits{R.} \bsnm{{Zhou}}},
\oauthor{\binits{A.} \bsnm{{Zitrin}}},
\oauthor{\binits{R.J.} \bsnm{{Foley}}},
\oauthor{\binits{S.W.} \bsnm{{Jha}}},
\oauthor{\binits{N.} \bsnm{{Kaiser}}},
\oauthor{\binits{V.} \bsnm{{Mehta}}},
\oauthor{\binits{S.} \bsnm{{Rieck}}},
\oauthor{\binits{L.} \bsnm{{Salo}}},
\oauthor{\binits{N.} \bsnm{{Smith}}},
\oauthor{\binits{D.R.} \bsnm{{Weisz}}},
{Flashlights: More than A Dozen High-Significance Microlensing Events of
  Extremely Magnified Stars in Galaxies at Redshifts z=0.7-1.5}.
arXiv e-prints,
2211--02670
(2022).
\doiurl{https://doi.org/10.48550/arXiv.2211.02670}
\end{botherref}
\endbibitem

\bibitem[\protect\citeauthoryear{{Mediavilla}
  et~al.}{2017}]{2017ApJ...836L..18M}
\begin{barticle}
\bauthor{\binits{E.} \bsnm{{Mediavilla}}},
\bauthor{\binits{J.} \bsnm{{Jim{\'e}nez-Vicente}}},
\bauthor{\binits{J.A.} \bsnm{{Mu{\~n}oz}}},
\bauthor{\binits{H.} \bsnm{{Vives-Arias}}},
\bauthor{\binits{J.} \bsnm{{Calder{\'o}n-Infante}}},
\batitle{{Limits on the Mass and Abundance of Primordial Black Holes from
  Quasar Gravitational Microlensing}}.
\bjtitle{\apjl}
\bvolume{836}(\bissue{2}),
\bfpage{18}
(\byear{2017}).
\doiurl{https://doi.org/10.3847/2041-8213/aa5dab}
\end{barticle}
\endbibitem

\bibitem[\protect\citeauthoryear{{Meena} et~al.}{2023a}]{MCZ23}
\begin{barticle}
\bauthor{\binits{A.K.} \bsnm{{Meena}}},
\bauthor{\binits{W.} \bsnm{{Chen}}},
\bauthor{\binits{A.} \bsnm{{Zitrin}}},
\bauthor{\binits{P.L.} \bsnm{{Kelly}}},
\bauthor{\binits{M.} \bsnm{{Golubchik}}},
\bauthor{\binits{R.} \bsnm{{Zhou}}},
\bauthor{\binits{A.} \bsnm{{Alfred}}},
\bauthor{\binits{T.} \bsnm{{Broadhurst}}},
\bauthor{\binits{J.M.} \bsnm{{Diego}}},
\bauthor{\binits{A.V.} \bsnm{{Filippenko}}},
\bauthor{\binits{S.K.} \bsnm{{Li}}},
\bauthor{\binits{M.} \bsnm{{Oguri}}},
\bauthor{\binits{N.} \bsnm{{Smith}}},
\bauthor{\binits{L.L.R.} \bsnm{{Williams}}},
\batitle{{Flashlights: an off-caustic lensed star at redshift z = 1.26 in Abell
  370}}.
\bjtitle{\mnras}
\bvolume{521}(\bissue{4}),
\bfpage{5224}--\blpage{5231}
(\byear{2023}a).
\doiurl{https://doi.org/10.1093/mnras/stad869}
\end{barticle}
\endbibitem

\bibitem[\protect\citeauthoryear{{Meena} et~al.}{2023b}]{MZJ23}
\begin{barticle}
\bauthor{\binits{A.K.} \bsnm{{Meena}}},
\bauthor{\binits{A.} \bsnm{{Zitrin}}},
\bauthor{\binits{Y.} \bsnm{{Jim{\'e}nez-Teja}}},
\bauthor{\binits{E.} \bsnm{{Zackrisson}}},
\bauthor{\binits{W.} \bsnm{{Chen}}},
\bauthor{\binits{D.} \bsnm{{Coe}}},
\bauthor{\binits{J.M.} \bsnm{{Diego}}},
\bauthor{\binits{P.} \bsnm{{Dimauro}}},
\bauthor{\binits{L.J.} \bsnm{{Furtak}}},
\bauthor{\binits{P.L.} \bsnm{{Kelly}}},
\bauthor{\binits{M.} \bsnm{{Oguri}}},
\bauthor{\binits{B.} \bsnm{{Welch}}},
\bauthor{\bsnm{{Abdurro'uf}}},
\bauthor{\binits{F.} \bsnm{{Andrade-Santos}}},
\bauthor{\binits{A.} \bsnm{{Adamo}}},
\bauthor{\binits{R.} \bsnm{{Bhatawdekar}}},
\bauthor{\binits{M.} \bsnm{{Brada{\v{c}}}}},
\bauthor{\binits{L.D.} \bsnm{{Bradley}}},
\bauthor{\binits{T.} \bsnm{{Broadhurst}}},
\bauthor{\binits{C.J.} \bsnm{{Conselice}}},
\bauthor{\binits{P.} \bsnm{{Dayal}}},
\bauthor{\binits{M.} \bsnm{{Donahue}}},
\bauthor{\binits{B.L.} \bsnm{{Frye}}},
\bauthor{\binits{S.} \bsnm{{Fujimoto}}},
\bauthor{\binits{T.Y.-Y.} \bsnm{{Hsiao}}},
\bauthor{\binits{V.} \bsnm{{Kokorev}}},
\bauthor{\binits{G.} \bsnm{{Mahler}}},
\bauthor{\binits{E.} \bsnm{{Vanzella}}},
\bauthor{\binits{R.A.} \bsnm{{Windhorst}}},
\batitle{{Two Lensed Star Candidates at z = 4.8 behind the Galaxy Cluster MACS
  J0647.7+7015}}.
\bjtitle{\apjl}
\bvolume{944}(\bissue{1}),
\bfpage{6}
(\byear{2023}b).
\doiurl{https://doi.org/10.3847/2041-8213/acb645}
\end{barticle}
\endbibitem

\bibitem[\protect\citeauthoryear{{Miralda-Escud\'e}}{1991}]{1991ApJ...379...94M}
\begin{barticle}
\bauthor{\binits{J.} \bsnm{{Miralda-Escud\'e}}},
\batitle{{The Magnification of Stars Crossing a Caustic. I. Lenses with Smooth
  Potentials}}.
\bjtitle{\apj}
\bvolume{379},
\bfpage{94}
(\byear{1991}).
\doiurl{https://doi.org/10.1086/170486}
\end{barticle}
\endbibitem

\bibitem[\protect\citeauthoryear{{Neindorf}}{2003}]{2003A&A...404...83N}
\begin{barticle}
\bauthor{\binits{B.} \bsnm{{Neindorf}}},
\batitle{{A probability theoretical access to extragalactic microlensing}}.
\bjtitle{\aap}
\bvolume{404},
\bfpage{83}--\blpage{92}
(\byear{2003}).
\doiurl{https://doi.org/10.1051/0004-6361:20030098}
\end{barticle}
\endbibitem

\bibitem[\protect\citeauthoryear{{Oguri}}{2010}]{2010PASJ...62.1017O}
\begin{barticle}
\bauthor{\binits{M.} \bsnm{{Oguri}}},
\batitle{{The Mass Distribution of SDSS J1004+4112 Revisited}}.
\bjtitle{PASJ}
\bvolume{62},
\bfpage{1017}
(\byear{2010}).
\doiurl{https://doi.org/10.1093/pasj/62.4.1017}
\end{barticle}
\endbibitem

\bibitem[\protect\citeauthoryear{{Oguri} et~al.}{2018}]{2018PhRvD..97b3518O}
\begin{barticle}
\bauthor{\binits{M.} \bsnm{{Oguri}}},
\bauthor{\binits{J.M.} \bsnm{{Diego}}},
\bauthor{\binits{N.} \bsnm{{Kaiser}}},
\bauthor{\binits{P.L.} \bsnm{{Kelly}}},
\bauthor{\binits{T.} \bsnm{{Broadhurst}}},
\batitle{{Understanding caustic crossings in giant arcs: Characteristic scales,
  event rates, and constraints on compact dark matter}}.
\bjtitle{\prd}
\bvolume{97}(\bissue{2}),
\bfpage{023518}
(\byear{2018}).
\doiurl{https://doi.org/10.1103/PhysRevD.97.023518}
\end{barticle}
\endbibitem

\bibitem[\protect\citeauthoryear{{Paczynski}}{1986}]{1986ApJ...301..503P}
\begin{barticle}
\bauthor{\binits{B.} \bsnm{{Paczynski}}},
\batitle{{Gravitational Microlensing at Large Optical Depth}}.
\bjtitle{\apj}
\bvolume{301},
\bfpage{503}
(\byear{1986}).
\doiurl{https://doi.org/10.1086/163919}
\end{barticle}
\endbibitem

\bibitem[\protect\citeauthoryear{{Petters} et~al.}{2009}]{2009JMP....50l2501P}
\begin{barticle}
\bauthor{\binits{A.O.} \bsnm{{Petters}}},
\bauthor{\binits{B.} \bsnm{{Rider}}},
\bauthor{\binits{A.M.} \bsnm{{Teguia}}},
\batitle{{A mathematical theory of stochastic microlensing. II. Random images,
  shear, and the Kac-Rice formula}}.
\bjtitle{Journal of Mathematical Physics}
\bvolume{50}(\bissue{12}),
\bfpage{122501}--\blpage{122501}
(\byear{2009}).
\doiurl{https://doi.org/10.1063/1.3267859}
\end{barticle}
\endbibitem

\bibitem[\protect\citeauthoryear{{Rodney} et~al.}{2018}]{2018NatAs...2..324R}
\begin{barticle}
\bauthor{\binits{S.A.} \bsnm{{Rodney}}},
\bauthor{\binits{I.} \bsnm{{Balestra}}},
\bauthor{\binits{M.} \bsnm{{Bradac}}},
\bauthor{\binits{G.} \bsnm{{Brammer}}},
\bauthor{\binits{T.} \bsnm{{Broadhurst}}},
\bauthor{\binits{G.B.} \bsnm{{Caminha}}},
\bauthor{\binits{G.} \bsnm{{Chiriv{\i}}}},
\bauthor{\binits{J.M.} \bsnm{{Diego}}},
\bauthor{\binits{A.V.} \bsnm{{Filippenko}}},
\bauthor{\binits{R.J.} \bsnm{{Foley}}},
\bauthor{\binits{O.} \bsnm{{Graur}}},
\bauthor{\binits{C.} \bsnm{{Grillo}}},
\bauthor{\binits{S.} \bsnm{{Hemmati}}},
\bauthor{\binits{J.} \bsnm{{Hjorth}}},
\bauthor{\binits{A.} \bsnm{{Hoag}}},
\bauthor{\binits{M.} \bsnm{{Jauzac}}},
\bauthor{\binits{S.W.} \bsnm{{Jha}}},
\bauthor{\binits{R.} \bsnm{{Kawamata}}},
\bauthor{\binits{P.L.} \bsnm{{Kelly}}},
\bauthor{\binits{C.} \bsnm{{McCully}}},
\bauthor{\binits{B.} \bsnm{{Mobasher}}},
\bauthor{\binits{A.} \bsnm{{Molino}}},
\bauthor{\binits{M.} \bsnm{{Oguri}}},
\bauthor{\binits{J.} \bsnm{{Richard}}},
\bauthor{\binits{A.G.} \bsnm{{Riess}}},
\bauthor{\binits{P.} \bsnm{{Rosati}}},
\bauthor{\binits{K.B.} \bsnm{{Schmidt}}},
\bauthor{\binits{J.} \bsnm{{Selsing}}},
\bauthor{\binits{K.} \bsnm{{Sharon}}},
\bauthor{\binits{L.-G.} \bsnm{{Strolger}}},
\bauthor{\binits{S.H.} \bsnm{{Suyu}}},
\bauthor{\binits{T.} \bsnm{{Treu}}},
\bauthor{\binits{B.J.} \bsnm{{Weiner}}},
\bauthor{\binits{L.L.R.} \bsnm{{Williams}}},
\bauthor{\binits{A.} \bsnm{{Zitrin}}},
\batitle{{Two peculiar fast transients in a strongly lensed host galaxy}}.
\bjtitle{Nature Astronomy}
\bvolume{2},
\bfpage{324}--\blpage{333}
(\byear{2018}).
\doiurl{https://doi.org/10.1038/s41550-018-0405-4}
\end{barticle}
\endbibitem

\bibitem[\protect\citeauthoryear{{Saha} and
  {Williams}}{2011}]{2011MNRAS.411.1671S}
\begin{barticle}
\bauthor{\binits{P.} \bsnm{{Saha}}},
\bauthor{\binits{L.L.R.} \bsnm{{Williams}}},
\batitle{{Understanding micro-image configurations in quasar microlensing}}.
\bjtitle{\mnras}
\bvolume{411}(\bissue{3}),
\bfpage{1671}--\blpage{1677}
(\byear{2011}).
\doiurl{https://doi.org/10.1111/j.1365-2966.2010.17797.x}
\end{barticle}
\endbibitem

\bibitem[\protect\citeauthoryear{{Sasaki} et~al.}{2018}]{2018CQGra..35f3001S}
\begin{barticle}
\bauthor{\binits{M.} \bsnm{{Sasaki}}},
\bauthor{\binits{T.} \bsnm{{Suyama}}},
\bauthor{\binits{T.} \bsnm{{Tanaka}}},
\bauthor{\binits{S.} \bsnm{{Yokoyama}}},
\batitle{{Primordial black holes{\textemdash}perspectives in gravitational wave
  astronomy}}.
\bjtitle{Classical and Quantum Gravity}
\bvolume{35}(\bissue{6}),
\bfpage{063001}
(\byear{2018}).
\doiurl{https://doi.org/10.1088/1361-6382/aaa7b4}
\end{barticle}
\endbibitem

\bibitem[\protect\citeauthoryear{{Schechter} and
  {Wambsganss}}{2002}]{2002ApJ...580..685S}
\begin{barticle}
\bauthor{\binits{P.L.} \bsnm{{Schechter}}},
\bauthor{\binits{J.} \bsnm{{Wambsganss}}},
\batitle{{Quasar Microlensing at High Magnification and the Role of Dark
  Matter: Enhanced Fluctuations and Suppressed Saddle Points}}.
\bjtitle{\apj}
\bvolume{580}(\bissue{2}),
\bfpage{685}--\blpage{695}
(\byear{2002}).
\doiurl{https://doi.org/10.1086/343856}
\end{barticle}
\endbibitem

\bibitem[\protect\citeauthoryear{{Schneider}
  et~al.}{1992}]{1992grle.book.....S}
\begin{bbook}
\bauthor{\binits{P.} \bsnm{{Schneider}}},
\bauthor{\binits{J.} \bsnm{{Ehlers}}},
\bauthor{\binits{E.E.} \bsnm{{Falco}}},
\bbtitle{{Gravitational Lenses}}
\byear{1992}.
\doiurl{https://doi.org/10.1007/978-3-662-03758-4}
\end{bbook}
\endbibitem

\bibitem[\protect\citeauthoryear{{Vall M{\"u}ller} and
  {Miralda-Escud{\'e}}}{2024}]{2024arXiv240316989V}
\begin{botherref}
\oauthor{\binits{C.} \bsnm{{Vall M{\"u}ller}}},
\oauthor{\binits{J.} \bsnm{{Miralda-Escud{\'e}}}},
{Limits on Dark Matter Compact Objects implied by Supermagnified Stars in
  Lensing Clusters}.
arXiv e-prints,
2403--16989
(2024).
\doiurl{https://doi.org/10.48550/arXiv.2403.16989}
\end{botherref}
\endbibitem

\bibitem[\protect\citeauthoryear{{Venumadhav}
  et~al.}{2017}]{2017ApJ...850...49V}
\begin{barticle}
\bauthor{\binits{T.} \bsnm{{Venumadhav}}},
\bauthor{\binits{L.} \bsnm{{Dai}}},
\bauthor{\binits{J.} \bsnm{{Miralda-Escud{\'e}}}},
\batitle{{Microlensing of Extremely Magnified Stars near Caustics of Galaxy
  Clusters}}.
\bjtitle{\apj}
\bvolume{850}(\bissue{1}),
\bfpage{49}
(\byear{2017}).
\doiurl{https://doi.org/10.3847/1538-4357/aa9575}
\end{barticle}
\endbibitem

\bibitem[\protect\citeauthoryear{{Vernardos}
  et~al.}{2024}]{2024SSRv..220...14V}
\begin{barticle}
\bauthor{\binits{G.} \bsnm{{Vernardos}}},
\bauthor{\binits{D.} \bsnm{{Sluse}}},
\bauthor{\binits{D.} \bsnm{{Pooley}}},
\bauthor{\binits{R.W.} \bsnm{{Schmidt}}},
\bauthor{\binits{M.} \bsnm{{Millon}}},
\bauthor{\binits{L.} \bsnm{{Weisenbach}}},
\bauthor{\binits{V.} \bsnm{{Motta}}},
\bauthor{\binits{T.} \bsnm{{Anguita}}},
\bauthor{\binits{P.} \bsnm{{Saha}}},
\bauthor{\binits{M.} \bsnm{{O'Dowd}}},
\bauthor{\binits{A.} \bsnm{{Peel}}},
\bauthor{\binits{P.L.} \bsnm{{Schechter}}},
\batitle{{Microlensing of Strongly Lensed Quasars}}.
\bjtitle{\ssr}
\bvolume{220}(\bissue{1}),
\bfpage{14}
(\byear{2024}).
\doiurl{https://doi.org/10.1007/s11214-024-01043-8}
\end{barticle}
\endbibitem

\bibitem[\protect\citeauthoryear{{Wambsganss}}{1990}]{1990PhDT.......180W}
\begin{botherref}
\oauthor{\binits{J.} \bsnm{{Wambsganss}}},
PhD thesis,
-,
1990
\end{botherref}
\endbibitem

\bibitem[\protect\citeauthoryear{{Wambsganss}}{1999}]{1999JCoAM.109..353W}
\begin{barticle}
\bauthor{\binits{J.} \bsnm{{Wambsganss}}},
\batitle{{Gravitational lensing: numerical simulations with a hierarchical tree
  code.}}
\bjtitle{Journal of Computational and Applied Mathematics}
\bvolume{109}(\bissue{1}),
\bfpage{353}--\blpage{372}
(\byear{1999})
\end{barticle}
\endbibitem

\bibitem[\protect\citeauthoryear{{Welch} et~al.}{2022a}]{WCD22}
\begin{barticle}
\bauthor{\binits{B.} \bsnm{{Welch}}},
\bauthor{\binits{D.} \bsnm{{Coe}}},
\bauthor{\binits{J.M.} \bsnm{{Diego}}},
\bauthor{\binits{A.} \bsnm{{Zitrin}}},
\bauthor{\binits{E.} \bsnm{{Zackrisson}}},
\bauthor{\binits{P.} \bsnm{{Dimauro}}},
\bauthor{\binits{Y.} \bsnm{{Jim{\'e}nez-Teja}}},
\bauthor{\binits{P.} \bsnm{{Kelly}}},
\bauthor{\binits{G.} \bsnm{{Mahler}}},
\bauthor{\binits{M.} \bsnm{{Oguri}}},
\bauthor{\binits{F.X.} \bsnm{{Timmes}}},
\bauthor{\binits{R.} \bsnm{{Windhorst}}},
\bauthor{\binits{M.} \bsnm{{Florian}}},
\bauthor{\binits{S.E.} \bsnm{{de Mink}}},
\bauthor{\binits{R.J.} \bsnm{{Avila}}},
\bauthor{\binits{J.} \bsnm{{Anderson}}},
\bauthor{\binits{L.} \bsnm{{Bradley}}},
\bauthor{\binits{K.} \bsnm{{Sharon}}},
\bauthor{\binits{A.} \bsnm{{Vikaeus}}},
\bauthor{\binits{S.} \bsnm{{McCandliss}}},
\bauthor{\binits{M.} \bsnm{{Brada{\v{c}}}}},
\bauthor{\binits{J.} \bsnm{{Rigby}}},
\bauthor{\binits{B.} \bsnm{{Frye}}},
\bauthor{\binits{S.} \bsnm{{Toft}}},
\bauthor{\binits{V.} \bsnm{{Strait}}},
\bauthor{\binits{M.} \bsnm{{Trenti}}},
\bauthor{\binits{S.} \bsnm{{Sharma}}},
\bauthor{\binits{F.} \bsnm{{Andrade-Santos}}},
\bauthor{\binits{T.} \bsnm{{Broadhurst}}},
\batitle{{A highly magnified star at redshift 6.2}}.
\bjtitle{\nat}
\bvolume{603}(\bissue{7903}),
\bfpage{815}--\blpage{818}
(\byear{2022}a).
\doiurl{https://doi.org/10.1038/s41586-022-04449-y}
\end{barticle}
\endbibitem

\bibitem[\protect\citeauthoryear{{Welch} et~al.}{2022b}]{WCZ22}
\begin{barticle}
\bauthor{\binits{B.} \bsnm{{Welch}}},
\bauthor{\binits{D.} \bsnm{{Coe}}},
\bauthor{\binits{E.} \bsnm{{Zackrisson}}},
\bauthor{\binits{S.E.} \bsnm{{de Mink}}},
\bauthor{\binits{S.} \bsnm{{Ravindranath}}},
\bauthor{\binits{J.} \bsnm{{Anderson}}},
\bauthor{\binits{G.} \bsnm{{Brammer}}},
\bauthor{\binits{L.} \bsnm{{Bradley}}},
\bauthor{\binits{J.} \bsnm{{Yoon}}},
\bauthor{\binits{P.} \bsnm{{Kelly}}},
\bauthor{\binits{J.M.} \bsnm{{Diego}}},
\bauthor{\binits{R.} \bsnm{{Windhorst}}},
\bauthor{\binits{A.} \bsnm{{Zitrin}}},
\bauthor{\binits{P.} \bsnm{{Dimauro}}},
\bauthor{\binits{Y.} \bsnm{{Jim{\'e}nez-Teja}}},
\bauthor{\bsnm{{Abdurro'uf}}},
\bauthor{\binits{M.} \bsnm{{Nonino}}},
\bauthor{\binits{A.} \bsnm{{Acebron}}},
\bauthor{\binits{F.} \bsnm{{Andrade-Santos}}},
\bauthor{\binits{R.J.} \bsnm{{Avila}}},
\bauthor{\binits{M.B.} \bsnm{{Bayliss}}},
\bauthor{\binits{A.} \bsnm{{Ben{\'\i}tez}}},
\bauthor{\binits{T.} \bsnm{{Broadhurst}}},
\bauthor{\binits{R.} \bsnm{{Bhatawdekar}}},
\bauthor{\binits{M.} \bsnm{{Brada{\v{c}}}}},
\bauthor{\binits{G.B.} \bsnm{{Caminha}}},
\bauthor{\binits{W.} \bsnm{{Chen}}},
\bauthor{\binits{J.} \bsnm{{Eldridge}}},
\bauthor{\binits{E.} \bsnm{{Farag}}},
\bauthor{\binits{M.} \bsnm{{Florian}}},
\bauthor{\binits{B.} \bsnm{{Frye}}},
\bauthor{\binits{S.} \bsnm{{Fujimoto}}},
\bauthor{\binits{S.} \bsnm{{Gomez}}},
\bauthor{\binits{A.} \bsnm{{Henry}}},
\bauthor{\binits{T.Y.-Y.} \bsnm{{Hsiao}}},
\bauthor{\binits{T.A.} \bsnm{{Hutchison}}},
\bauthor{\binits{B.L.} \bsnm{{James}}},
\bauthor{\binits{M.} \bsnm{{Joyce}}},
\bauthor{\binits{I.} \bsnm{{Jung}}},
\bauthor{\binits{G.} \bsnm{{Khullar}}},
\bauthor{\binits{R.L.} \bsnm{{Larson}}},
\bauthor{\binits{G.} \bsnm{{Mahler}}},
\bauthor{\binits{N.} \bsnm{{Mandelker}}},
\bauthor{\binits{S.} \bsnm{{McCandliss}}},
\bauthor{\binits{T.} \bsnm{{Morishita}}},
\bauthor{\binits{R.} \bsnm{{Newshore}}},
\bauthor{\binits{C.} \bsnm{{Norman}}},
\bauthor{\binits{K.} \bsnm{{O'Connor}}},
\bauthor{\binits{P.A.} \bsnm{{Oesch}}},
\bauthor{\binits{M.} \bsnm{{Oguri}}},
\bauthor{\binits{M.} \bsnm{{Ouchi}}},
\bauthor{\binits{M.} \bsnm{{Postman}}},
\bauthor{\binits{J.R.} \bsnm{{Rigby}}},
\bauthor{\binits{J.} \bsnm{{Ryan}} \bsuffix{Russell~E.}},
\bauthor{\binits{S.} \bsnm{{Sharma}}},
\bauthor{\binits{K.} \bsnm{{Sharon}}},
\bauthor{\binits{V.} \bsnm{{Strait}}},
\bauthor{\binits{L.-G.} \bsnm{{Strolger}}},
\bauthor{\binits{F.X.} \bsnm{{Timmes}}},
\bauthor{\binits{S.} \bsnm{{Toft}}},
\bauthor{\binits{M.} \bsnm{{Trenti}}},
\bauthor{\binits{E.} \bsnm{{Vanzella}}},
\bauthor{\binits{A.} \bsnm{{Vikaeus}}},
\batitle{{JWST Imaging of Earendel, the Extremely Magnified Star at Redshift z
  = 6.2}}.
\bjtitle{\apjl}
\bvolume{940}(\bissue{1}),
\bfpage{1}
(\byear{2022}b).
\doiurl{https://doi.org/10.3847/2041-8213/ac9d39}
\end{barticle}
\endbibitem

\bibitem[\protect\citeauthoryear{{Windhorst}
  et~al.}{2018}]{2018ApJS..234...41W}
\begin{barticle}
\bauthor{\binits{R.A.} \bsnm{{Windhorst}}},
\bauthor{\binits{F.X.} \bsnm{{Timmes}}},
\bauthor{\binits{J.S.B.} \bsnm{{Wyithe}}},
\bauthor{\binits{M.} \bsnm{{Alpaslan}}},
\bauthor{\binits{S.K.} \bsnm{{Andrews}}},
\bauthor{\binits{D.} \bsnm{{Coe}}},
\bauthor{\binits{J.M.} \bsnm{{Diego}}},
\bauthor{\binits{M.} \bsnm{{Dijkstra}}},
\bauthor{\binits{S.P.} \bsnm{{Driver}}},
\bauthor{\binits{P.L.} \bsnm{{Kelly}}},
\bauthor{\binits{D.} \bsnm{{Kim}}},
\batitle{{On the Observability of Individual Population III Stars and Their
  Stellar-mass Black Hole Accretion Disks through Cluster Caustic Transits}}.
\bjtitle{\apjs}
\bvolume{234}(\bissue{2}),
\bfpage{41}
(\byear{2018}).
\doiurl{https://doi.org/10.3847/1538-4365/aaa760}
\end{barticle}
\endbibitem

\end{thebibliography}
\nocite{*}

\end{document}